\documentclass[a4paper,fleqn]{cas-sc}

\usepackage[authoryear]{natbib}
\setcitestyle{authoryear, maxcitenames=2, maxbibnames=2}
\usepackage{tcolorbox}
\tcbuselibrary{listingsutf8}
\newtcolorbox[auto counter]{promptbox}[2][]{%
  colback=gray!5!white, 
  colframe=blue!60!teal, 
  fonttitle=\bfseries,
  fontupper=\footnotesize,
  title=Prompt~\thetcbcounter: #2, 
  label={#1}
}
\definecolor{RowGray}{gray}{0.95}
\usepackage{subcaption}
\usepackage{longtable}
\usepackage[table]{xcolor}
\usepackage{array}
\usepackage{tabularx}
\usepackage{xcolor,colortbl}
\usepackage{float}
\usepackage{stfloats}          % improved two-column *-float placement
\usepackage{array}
\usepackage{booktabs}
\graphicspath{{Figures/}}

\usepackage{acro}
\acsetup{
  first-style = long-short,
}
\DeclareAcronym{AI}{short=AI, long=artificial intelligence}
\DeclareAcronym{AV}{short=AV, long=autonomous vehicle}
\DeclareAcronym{PCSC}{short=PCSC, long=Perceived Community Driving-Safety Concern}
\DeclareAcronym{GAAIS}{short=GAAIS, long=General Attitudes towards Artificial Intelligence Scale}
\DeclareAcronym{TAM}{short=TAM, long=Technology Acceptance Model}
\DeclareAcronym{UTAUT}{short=UTAUT, long=Unified Theory of Acceptance and Use of Technology}
\DeclareAcronym{PPM}{short=PPM, long=Push-Pull-Mooring}
\DeclareAcronym{ATP}{short=ATP, long=American Trends Panel}
\DeclareAcronym{SEM}{short=SEM, long=structural equation modeling}
\DeclareAcronym{CFA}{short=CFA, long=confirmatory factor analysis, short-plural-form=CFA}
\DeclareAcronym{WLSMV}{short=WLSMV, long=Weighted Least Squares Mean and Variance Adjusted}
\DeclareAcronym{HTMT}{short=HTMT, long=Heterotrait-Monotrait ratio of correlations}
\DeclareAcronym{AVE}{short=AVE, long=average variance extracted}
\DeclareAcronym{PSM}{short=PSM, long=propensity score matching}
\DeclareAcronym{MAR}{short=MAR, long=missing at random}
\DeclareAcronym{BC}{short=BC, long=bias-corrected}
\DeclareAcronym{CFI}{short=CFI, long=Comparative Fit Index}
\DeclareAcronym{TLI}{short=TLI, long=Tucker-Lewis Index}
\DeclareAcronym{RMSEA}{short=RMSEA, long=root mean square error of approximation}
\DeclareAcronym{SRMR}{short=SRMR, long=standardized root mean residual}
\usepackage{xcolor}
\usepackage{longtable,booktabs,array}
\usepackage{multirow}
\usepackage{calc} % for calculating minipage widths
\usepackage{etoolbox}
\makeatletter
\patchcmd\longtable{\par}{\if@noskipsec\mbox{}\fi\par}{}{}
\makeatother
\IfFileExists{footnotehyper.sty}{\usepackage{footnotehyper}}{\usepackage{footnote}}
\makesavenoteenv{longtable}
\ifLuaTeX
  \usepackage{selnolig}  % disable illegal ligatures
\fi
\IfFileExists{bookmark.sty}{\usepackage{bookmark}}{\usepackage{hyperref}}
\IfFileExists{xurl.sty}{\usepackage{xurl}}{} % add URL line breaks if available
\hypersetup{
  pdftitle={Community Driving-Safety Deterioration as a Push Factor for Public Endorsement of AI Driving Capability},
  hidelinks,
  pdfcreator={LaTeX via pandoc}}
\def\tsc#1{\csdef{#1}{\textsc{\lowercase{#1}}\xspace}}
\tsc{WGM}
\tsc{QE}
\tsc{EP}
\tsc{PMS}
\tsc{BEC}
\tsc{DE}
\begin{document}
\let\WriteBookmarks\relax
% Two-column star floats (table*/figure*): avoid piling everything at the top and
% reduce vertical stretch gaps from deferred floats (see also \clearpage before back matter).
\setcounter{dbltopnumber}{4}
\renewcommand{\dbltopfraction}{0.85}
\renewcommand{\dblfloatpagefraction}{0.75}
\raggedbottom

% Short title
\shorttitle{Community Driving-Safety Deterioration and AI Driving Capability}

% Short author
\shortauthors{Rafe et~al.}

% Main title of the paper
\title [mode = title]{Community Driving-Safety Deterioration as a Push Factor for Public Endorsement of AI Driving Capability}     

% First author
%
% Options: Use if required
% eg: \author[1,3]{Author Name}[type=editor,
%       style=chinese,
%       auid=000,
%       bioid=1,
%       prefix=Sir,
%       orcid=0000-0000-0000-0000,
%       facebook=<facebook id>,
%       twitter=<twitter id>,
%       linkedin=<linkedin id>,
%       gplus=<gplus id>]
\author[1]{Amir Rafe}[orcid=0000-0002-4089-2088]
\cormark[1]
\ead{amir.rafe@txstate.edu}
\credit{Conceptualization, Methodology, Software, Formal analysis, Writing -- original draft, Visualization}

\author[1]{Subasish Das}[orcid=0000-0002-1671-2753]
\ead{subasish@txstate.edu}
\credit{Conceptualization, Methodology, Supervision, Writing -- review \& editing}

% Address/affiliation
\affiliation[1]{organization={Civil Engineering, Texas State University},
    addressline={601 University Drive}, 
    city={San Marcos},
    % citysep={}, % Uncomment if no comma needed between city and postcode
    postcode={78666 TX}, 
    % state={},
    country={USA}}

% Corresponding author text
\cortext[cor1]{Corresponding author}

% Here goes the abstract
\begin{abstract}
Road traffic crashes claim approximately 1.19 million lives annually worldwide, and human error accounts for the vast majority, yet the autonomous vehicle acceptance literature models adoption almost exclusively through technology-centered pull factors such as perceived usefulness and trust.  This study examines a moderated mediation model in which perceived community driving-safety concern (PCSC) predicts evaluations of AI versus human driving capability, mediated by Generalized AI Orientation and moderated by personal driving frequency. Weighted structural equation modeling is applied to a nationally representative U.S. probability sample from Pew Research Center's American Trends Panel Wave~152, using Weighted Least Squares Mean and Variance Adjusted (WLSMV)-estimated confirmatory factor analysis on ordinal indicators, bias-corrected bootstrap inference, and seven robustness checks including Imai sensitivity analysis, E-value confounding thresholds, and propensity score matching. Results reveal a dual-pathway mechanism constituting an inconsistent mediation: PCSC exerts a small positive direct effect on AI driving evaluation, consistent with a domain-specific push interpretation, while simultaneously suppressing Generalized AI Orientation, which is itself a strong positive predictor of AI driving evaluation. Conditional indirect effects are negative and statistically significant at low, mean, and high levels of driving frequency. These findings establish a risk-spillover mechanism whereby community driving-safety concern promotes domain-specific AI endorsement yet suppresses domain-general AI enthusiasm, yielding a near-zero net total effect. 
\end{abstract}

% Use if graphical abstract is present
% \begin{graphicalabstract}
% \includegraphics{figs/grabs.pdf}
% \end{graphicalabstract}

% Keywords
% Each keyword is separated by \sep

\maketitle

\section{Introduction}

Road traffic crashes remain one of the most pressing global public health challenges, claiming approximately 1.19 million lives annually worldwide and ranking as the leading cause of death among children and young adults aged 5 to 29 \citep{WHO2023}. In the United States, traffic fatalities surged above 40,000 annually during 2021--2023 before declining to an estimated 36,640 in 2025, a level that merely returns the nation to its pre-pandemic 2019 baseline of approximately 36,100 deaths per year \citep{NHTSA2024, NHTSA2026proj}. Even at the projected 2025 level, roughly 100 people die on U.S. roads every day, and human error, encompassing speeding, distracted driving, impaired driving, and aggressive maneuvers, accounts for the vast majority of these crashes. At the same time, \ac{AI} has advanced rapidly in the transportation domain, with \ac{AV} systems now undergoing large-scale deployment trials in multiple metropolitan areas. This convergence of persistent human driving failure and accelerating AI capability has created an urgent imperative to understand the conditions under which the public endorses algorithmic alternatives to human drivers. Public opinion data reflect the salience of this tension: a 2024 Pew Research Center survey found that 49\% of Americans believe people in their community are driving less safely compared with five years ago, while 78\% identify cellphone distraction as a major local problem \citep{PewDriving2024}. Simultaneously, the AAA Foundation for Traffic Safety (AAAFTS) reports that 96\% of drivers express concern about dangerous driving behaviors in their area, and aggressive driving is perceived as very or extremely dangerous by 90\% of respondents \citep{AAA2024TSCI}. These figures suggest that public dissatisfaction with the incumbent transportation system, specifically the quality of human driving, may constitute a significant but underexamined antecedent of openness to AI-driven alternatives.

The academic literature on AV acceptance has been dominated by a few cluster of technology-centered theoretical frameworks. The \ac{TAM} \citep{Davis1989} and the \ac{UTAUT} \citep{Venkatesh2003, Venkatesh2012} posit that behavioral intention toward a new technology is primarily a function of cognitive evaluations of the technology's attributes, most notably perceived usefulness and perceived ease of use. In the AV context, researchers have extended these foundational models by integrating high-stakes constructs such as trust, perceived risk, and perceived safety \citep{Zhangetal2019, Manetal2020, Buckleyetal2018}. Meta-analytic synthesis across dozens of studies confirms that trust is the most significant predictor of AV acceptance, followed by perceived usefulness, while perceived ease of use contributes least \citep{Zhang2021meta}. Systematic reviews further document that this literature is predominantly technology-centered: acceptance is modeled as a function of how favorable users perceive the AI system to be, rather than how unsatisfactory they find the incumbent alternative \citep{Jingetal2020, Golbabaei2020, Naisehetal2025}. Within this paradigm, ``usefulness'' is typically operationalized in abstract or future-oriented terms, such as environmental sustainability, generalized traffic efficiency, or the novelty of hands-free travel \citep{Zhang2021meta, Acheampong2019}. While these constructs are valid predictors of adoption intention, they reflect a pull logic in which the attractive features of a novel system draw users toward it, leaving unexamined whether dissatisfaction with the status quo exerts independent pressure in the same direction.

Safety beliefs occupy a central position in the AV discourse, yet they are framed almost exclusively through the lens of AV-specific risk rather than through a relative comparison with human driving performance. Public acceptance depends critically on whether respondents believe automated systems are safer or less error-prone than human drivers \citep{Nees2019, Liu2020, Shariff2021}, and nationally representative evidence demonstrates that consumers prefer lower automation levels for themselves than for others, a pattern driven by self-enhancing assessments of one's own driving ability rather than by different assessments of AV capabilities \citep{AgarwalDeFreitas2024}. The dominant direction of inquiry centers on fear of the machine: studies examine perceived AV risk, the social amplification of risk following high-profile crashes, and the lack of trust in automation \citep{Naisehetal2025, Montoro2019}. This focus implicitly treats the human driver as a stable and unproblematic comparison baseline, even though aggregated safety data show that human error is responsible for the vast majority of traffic fatalities. What the literature has not examined is whether deteriorating perceptions of that baseline, specifically the sense that human drivers in one's own community are increasingly dangerous, distracted, or aggressive, function as a push factor that reframes the AI-versus-human comparison in favor of algorithmic alternatives. The \ac{PPM} framework \citep{Bansal2005}, which has been applied to technology switching contexts including ride-hailing to robotaxi transitions \citep{Liu2025}, provides a theoretical vocabulary for this mechanism, yet no study has positioned perceived community driving-safety concern as the push element in a mediation model predicting AI driving evaluations.

Despite this gap, the empirical foundation for such a test exists. Large-scale traffic safety surveys consistently document extensive public concern about dangerous driving behaviors at the community level: the Pew Research Center reports that 78\% of Americans view cellphone distraction while driving as a major problem in their area, 63\% identify speeding as a major concern, and nearly one quarter report witnessing road rage often \citep{PewDriving2024}. The AAAFTS Traffic Safety Culture Index confirms that perceptions of dangerous driving vary systematically by driving frequency, urbanicity, and age, with frequent drivers and suburban residents expressing higher concern about worsening road conditions \citep{AAA2024TSCI, AAA2025RoadRage}. State-level data from California's annual traffic safety survey show that speeding and aggressive driving have ranked as the top safety concern for nine consecutive years, a pattern that intensified following the COVID-19 pandemic \citep{CaliforniaTSS2025}. However, these indicators have been treated almost exclusively as descriptive benchmarks for public opinion or as outcomes of traffic safety policy; they have not been modeled as predictors of technology preference. Concurrently, a growing body of evidence demonstrates that general attitudes toward artificial intelligence, encompassing excitement, concern, trust, and perceived benefits, shape domain-specific AI judgments in healthcare, education, and public decision-making \citep{Schepman2020, Schepman2022, PewAIexperts2025, Gallup2024AI}. The \ac{GAAIS} captures a two-factor structure of positive and negative orientations that predict behavioral intentions across AI applications \citep{Schepman2022}. Yet general AI orientation has been positioned in the existing literature as a direct predictor or background covariate, not as a mediating mechanism linking a real-world environmental stressor to a specific AI performance judgment. The present study addresses this compounded gap by hypothesizing that perceived community driving-safety concern activates a broader shift in general AI orientation, which in turn shapes the specific comparative evaluation of AI versus human driving capability.

Three research questions guide the current investigation. First, does a latent construct of \ac{PCSC}, indicated by perceived severity of speeding, aggressive driving, alcohol-impaired driving, marijuana-impaired driving, cellphone distraction, and cyclist/pedestrian risk, significantly predict evaluations of AI (relative to humans) for a driving task? Second, is this relationship mediated by the respondent's Generalized AI Orientation, operationalized as the excited-versus-concerned disposition toward AI in daily life? Third, does personal driving frequency moderate the PCSC-to-Generalized AI Orientation pathway, and if so, in what direction and magnitude? The theoretical framework integrates the affect heuristic \citep{Slovic2002}, which predicts that perceived community driving danger activates a domain-general negative risk schema that spills over into general AI attitudes; risk homeostasis theory \citep{Wilde1982}, which suggests that even reduced AI enthusiasm may remain sufficiently positive relative to a deteriorating human-driver baseline to sustain favorable driving-specific evaluations; and prospect theory \citep{Kahneman1979}, which posits that community safety losses loom larger than equivalent gains and thereby color technology attitudes through loss-aversion. This integrated framework accounts for the empirically observed tension between a positive direct effect of PCSC on AI driving evaluation and a negative indirect effect operating through suppressed general AI orientation.

The study offers four principal contributions. First, it provides a theoretical contribution by revealing a risk-spillover mechanism: perceived community driving-safety concern suppresses general AI orientation through a negative mediated pathway yet simultaneously predicts more favorable AI driving evaluations through a positive direct effect, a tension between domain-general risk aversion and domain-specific pragmatic endorsement that refines and extends both TAM-based and push-pull frameworks in the AV acceptance literature. Second, it offers a methodological contribution by demonstrating rigorous weighted structural equation modeling with ordinal indicators on a probability-based national sample (Pew \ac{ATP} Wave~152), including \ac{WLSMV}-estimated \aclp{CFA}, bias-corrected bootstrap indirect effects, Johnson-Neyman floodlight analysis, Imai sensitivity analysis \citep{Imai2010}, E-value confounding thresholds \citep{VanderWeele2017}, \Ac{PSM}, and multigroup invariance tests. Third, the study provides an empirical contribution: conditional indirect effects at low, mean, and high driving frequency are all statistically significant and negative, indicating a consistent mediating process irrespective of how often respondents drive. Fourth, the findings carry policy relevance, suggesting that interventions aimed at AV acceptance may need to address two distinct pathways: directly reducing community driving-safety concerns through enforcement and traffic calming, and improving general AI literacy and orientation to prevent risk-spillover from suppressing AI evaluations.

The remainder of this paper is organized as follows. Section 2 reviews the relevant literature on AV acceptance, community driving-safety perceptions, and general AI attitudes. Section 3 describes the data source, measurement model, and structural equation modeling approach. Section 4 presents the results, including the measurement models, structural paths, indirect effects, and robustness checks. Section 5 discusses the theoretical and policy implications, and Section 6 concludes with limitations and directions for future research.

\section{Literature Review}

\subsection{Technology acceptance frameworks in the autonomous vehicle (AV) context}

Studies applying TAM to autonomous vehicle (AV) acceptance have consistently confirmed that perceived usefulness and perceived ease of use predict adoption intention, with perceived usefulness serving as the stronger factor \citep{Buckleyetal2018, Nastjuk2020}. Subsequent extensions have introduced trust as a pivotal construct that frequently surpasses TAM's original predictors in explanatory power. \citet{Zhangetal2019} extended TAM with initial trust, perceived safety risk, and perceived privacy risk and found that initial trust, rather than perceived usefulness alone, was the most critical determinant of attitude toward Level 3 AVs. \citet{Manetal2020} replicated this pattern among Hong Kong drivers, reporting that the full trust-augmented model explained 68\% of the variance in intention to use AVs when system compatibility and system quality were also included. In parallel, applications of the UTAUT and UTAUT2 frameworks have broadened the predictor set to encompass performance expectancy, social influence, hedonic motivation, and facilitating conditions. \citet{Madigan2017} applied UTAUT to automated public transport systems in Greece and found that hedonic motivation was the strongest predictor of behavioral intention, while effort expectancy failed to reach significance, suggesting that the experiential appeal of the technology may matter more than its perceived simplicity. A large-scale cross-national study of 9,118 drivers across eight European countries similarly found that performance expectancy, hedonic motivation, and social influence predicted acceptance of Level 3 vehicles, though the magnitude and ranking of these effects varied substantially by country \citep{Nordhoff2020}. \citet{Zhang2020social} extended TAM with social influence and personality traits in a Chinese sample of 647 drivers and concluded that social influence and initial trust were the strongest contributors, while sensation-seeking personality traits played a secondary role. Intrinsic motivation, personality characteristics, and perceived control have also been shown to add incremental variance in trust-augmented models \citep{Hegner2019}. Across this body of work, each added construct remains firmly on the technology side of the equation: the central explanatory mechanism is whether the user finds the AI system trustworthy, enjoyable, or performant.

A critical consequence of this technology-centered orientation is that the concept of usefulness is almost universally operationalized as an abstract, future-oriented benefit rather than as a response to identifiable failures of the system the technology would replace. Studies evaluating perceived usefulness ask whether the respondent expects AVs to reduce travel time, improve mobility, or enhance environmental sustainability \citep{Acheampong2019, Nastjuk2020}, but none anchor usefulness in the perception that local human driving has become unacceptably dangerous or unreliable. Even field experiments that expose participants to real automated driving experiences continue to evaluate acceptance through technology-attribute constructs rather than through incumbent-system dissatisfaction \citep{Xu2018}. This asymmetry means that the pull logic of technology attraction is well characterized, but the push logic of incumbent-system failure, a mechanism central to service-switching theory \citep{Bansal2005} and already applied to ride-hailing-to-robotaxi transitions \citep{Liu2025}, has been absent from quantitative AV acceptance models.

\subsection{Safety comparison and the asymmetry of risk framing}

Safety beliefs constitute the most consequential cluster of predictors in the AV acceptance literature, yet the evidence base reveals a systematic directional asymmetry: safety research focuses overwhelmingly on how risky respondents perceive autonomous systems to be, rather than on how risky they find the human drivers around them. \citet{Nees2019} demonstrated that the majority of respondents rate their own driving ability as above average, producing a benchmark against which no automated system can credibly compete; the popular argument that ``AVs are safer than the average human driver'' fails precisely because most people do not identify with the average. \citet{Liu2020} extended this observation experimentally, showing that matched safety performance between AI and human drivers is far from sufficient for public acceptance and that respondents demand substantially greater safety from automated systems before they regard the technology as acceptable. \citet{Shariff2021} traced these extreme safety demands to deeper psychological mechanisms, including algorithm aversion, the asymmetric moral outrage triggered by machine-caused harm relative to human-caused harm, and the intuition that human agency in a crash is preferable to algorithmic agency even when the outcomes are identical. Perceptions also vary by road-user type: \citet{Hulse2018} found that pedestrians and cyclists, who are disproportionately exposed to the consequences of human driving errors, perceived AVs more favorably than did regular drivers, suggesting that the human-driver baseline is not evaluated uniformly across all stakeholders. Research on real-world road tests of AVs has further demonstrated that both affective reactions (emotional responses during the ride) and cognitive appraisals (rational performance assessments) contribute to acceptance formation, but neither pathway incorporates the perceived quality of surrounding human driving as an input variable \citep{Liu2019road}. Across these studies, the dependent variable is typically generic AV favorability or adoption intention, framed as an absolute judgment about the technology rather than a relative comparison with the human-driver baseline.

\subsection{Community driving-safety concern as a demand-side antecedent}

Traffic safety culture research has produced extensive documentation of widespread concern about dangerous driving behaviors at the local level, yet this body of evidence has remained disconnected from the technology acceptance literature. The AAA Foundation for Traffic Safety administers an annual Traffic Safety Culture Index through a probability-based panel weighted to represent the U.S. adult driving population, tracking perceptions of driving-risk behaviors over time and grouping respondents into behavioral risk profiles based on self-reported engagement in speeding, distracted driving, impaired driving, and aggressive maneuvers \citep{AAA2024TSCI}. These risk profiles reveal systematic variation in how respondents perceive danger on the road: drivers who report fewer risky behaviors of their own tend to express greater concern about the behaviors of others, while higher-risk drivers are more likely to normalize the same behaviors \citep{AAA2025RoadRage}. Pew ATP Wave 152, the data source for the present study, covers a complementary set of indicators, including perceptions of whether community driving has deteriorated over the past five years and assessments of specific hazards such as cellphone distraction, speeding, and impaired driving \citep{PewATP152}. Research on age-group variation in AV acceptance has found that older drivers, who frequently report greater concern about the driving behaviors of others, are simultaneously more cautious about AV technology \citep{Classen2023}, suggesting that the relationship between community safety perception and AV endorsement may not be straightforwardly positive and may depend on mediating psychological processes. Despite the richness of this evidence, the AV literature has not modeled community-level driving-safety concern as a predictor of technology preference. Traffic safety surveys treat perceived danger as an outcome of policy interest, while AV acceptance studies treat technology attributes as inputs.

\subsection{General AI orientation as a mediating mechanism}

The transition from a community-level environmental concern to a favorable evaluation of AI on a specific task requires a psychological mechanism that connects a domain-general stressor to a domain-specific technology judgment, and general AI orientation is the most plausible candidate for that connecting role. The GAAIS, originally developed by \citet{Schepman2020} and subsequently validated in a larger confirmatory study \citep{Schepman2022}, captures a two-factor structure in which a positive subscale (reflecting perceived utility, efficiency, and beneficial societal impact) and a negative subscale (reflecting fears about dehumanization, loss of control, and job displacement) independently predict behavioral intentions toward AI-based products and services. Confirmatory work has replicated this structure across multiple national contexts and shown that AI knowledge enhances the positive dimension, while perceived societal risk accentuates the negative one \citep{Schepman2022}. Broader public opinion research corroborates this pattern at the population level, documenting that the balance between excitement and concern about AI varies substantially by age, education, and prior AI exposure \citep{PewAIexperts2025, Gallup2024AI}. In the AV context, general technological optimism and trust in AI have been identified as significant predictors of acceptance \citep{Hegner2019, Nastjuk2020}. However, in every case of which the authors are aware, general AI orientation has been positioned as either a direct predictor of AV acceptance or a moderating background variable; it has not been modeled as a mediator linking a real-world environmental condition to a specific AI performance judgment.

\subsection{Methodological considerations and synthesis}

The methodological practices that characterize the AV acceptance literature constrain both the internal validity and the generalizability of its accumulated findings. The dominant sampling strategy involves convenience recruitment, student populations, or online opt-in panels that systematically over-represent younger, more educated, and more technology-experienced respondents relative to the national adult population \citep{Nordhoff2020, Hulse2018}. Hypothetical scenario designs, in which respondents imagine riding in or evaluating an AV they have never encountered, introduce unknown discrepancies between stated and revealed preferences. The treatment of ordinal Likert-scale items as continuous variables without latent-variable correction is widespread, and survey weights that adjust for differential selection probability and nonresponse are rarely applied in structural models. Analytic designs are typically limited to direct-effect regressions or single-mediator \ac{SEM}; moderated mediation, which tests whether the strength of an indirect effect varies across levels of a moderating variable, is seldom reported. Although several recent studies have moved toward more complex structural specifications \citep{Nastjuk2020, Zhang2020social}, nationally representative, probability-weighted evidence remains the exception rather than the norm. The present study capitalizes on Pew \ac{ATP} Wave~152, a probability-based national panel weighted by gender, race, ethnicity, partisan affiliation, education, and other demographic dimensions \citep{PewATP152}, to estimate a \acs{WLSMV}-based \acl{SEM} model that accommodates ordinal indicators, incorporates survey-design corrections, and uses bias-corrected bootstrap inference for conditional indirect effects.

The foregoing review identifies three interconnected gaps that the present study addresses simultaneously. First, the AV acceptance literature has been built almost exclusively around technology-centered predictors, principally trust, perceived usefulness, and perceived AV risk, while the possibility that perceived failure of the incumbent human-driving system functions as a demand-side push factor has not been empirically tested in national probability-based data. Second, safety comparison is framed asymmetrically: fear of autonomous systems is extensively researched, but the deterioration of public confidence in human drivers has not been modeled as an antecedent of favorable AI driving evaluations. Third, general AI orientation, despite consistent evidence that it predicts AI evaluations across multiple application domains, has not been positioned as a mediating mechanism linking an environmental stressor to a domain-specific performance judgment.

The present study addresses these gaps through four linked design decisions. First, it positions perceived community driving-safety concern as the primary demand-side antecedent, asking not what makes AVs attractive but what makes human driving intolerable enough to reconfigure attitudes toward AI alternatives, consistent with push-pull logic in which dissatisfaction with the incumbent system generates the motivational energy for evaluating alternatives \citep{Bansal2005, Liu2025}. Second, it uses a direct AI-versus-human performance comparison on the driving task as its outcome, repositioning the human-driver baseline from an implicit default to an explicit analytical object. Third, it models general AI orientation as a mediator linking a real-world environmental stressor to a domain-specific technology judgment through the affect heuristic's risk-spillover mechanism, a formulation implying that the same level of community driving concern may produce different AI driving evaluations depending on the respondent's general orientation toward AI. Fourth, it estimates these relationships within a weighted moderated-mediation model on Pew \ac{ATP} Wave~152, using \acs{WLSMV}-based \acl{SEM} with ordinal indicators, bias-corrected bootstrap inference, and survey-design corrections, an analytic specification that exceeds the methodological rigor of the convenience-sample, direct-effect designs that dominate the existing literature.

\section{Data and Method}

\subsection{Data source and sample}

This study draws on the Pew Research Center's \ac{ATP}, Wave~152, a probability-based online panel of U.S. adults fielded between August 12 and 18, 2024 \citep{PewATP152}. The ATP recruits respondents through address-based sampling of the U.S. Postal Service Computerized Delivery Sequence File and applies post-stratification raking weights calibrated to the Current Population Survey benchmarks for gender, age, education, race and ethnicity, Hispanic nativity and language, U.S. Census region, metropolitan status, volunteerism, voter registration, and partisan affiliation. The survey was administered in English and Spanish. The total analytic sample is $N$ = 5,410. Because post-stratification weights inflate the nominal sample size, the Kish \citeyearpar{Kish1965} effective sample size is computed as

\begin{equation}
n_{\text{eff}} = \frac{\left(\sum_{i=1}^{N} w_i\right)^{2}}{\sum_{i=1}^{N} w_i^{2}} = 3{,}993
\label{eq:kish}
\end{equation}

\noindent where $w_i$ denotes the survey weight for respondent $i$. All inferential statistics, including fit indices, standard errors, and bootstrap confidence intervals, are computed with respect to $n_{\text{eff}}$ rather than the raw sample size.

\subsection{Measures}

Five categories of variables were identified from the codebook: a latent focal predictor (PCSC), a latent mediator (Generalized AI Orientation), an observed ordinal outcome (AI Driving Evaluation), an observed ordinal moderator (Driving Frequency), and a vector of control covariates. Across all construct indicators, Refuse and Don't Know responses (coded 99) were set to missing prior to analysis. Table~\ref{tab:variables} provides a complete mapping of constructs, indicators, survey items, response scales, item direction, and standardized factor loadings; this subsection describes each construct's operationalization.

%\subsubsection{Perceived Community Driving-Safety Concern (PCSC)}

PCSC is operationalized as a latent construct indicated by six ordinal items (DRIVE2\_a--f) in which respondents rated the severity of six driving problems in their local community: speeding, aggressive driving (tailgating, weaving, running red lights), alcohol-impaired driving, marijuana-impaired driving, cellphone distraction, and endangering cyclists or pedestrians. Original response categories (1 = \textit{Major problem}, 2 = \textit{Minor problem}, 3 = \textit{Not a problem}) were reverse-coded so that higher scores indicate greater perceived concern (1 = \textit{Not a problem}, 3 = \textit{Major problem}).

%\subsubsection{Generalized AI Orientation}

Generalized AI Orientation captures the respondent's overall excited-versus-concerned disposition toward artificial intelligence and is indicated by seven items drawn from two survey sections. From Section C (AI overview): the excited/concerned valence item (CNCEXC), perceived AI impact on U.S. life over the next 20 years (AICHANGE), personal benefit versus harm expectation (PERSBENHRM), comfort with control over AI use (AICONTROL2), and willingness to trust AI for important personal decisions (TRSTAIPRS). From Section E (AI concerns): concern that AI makes biased or discriminatory decisions (AICONCERN\_a) and concern that AI enables online impersonation (AICONCERN\_b). All items were recoded so that higher values reflect a more favorable (pro-AI) orientation. Because indicators span different response formats (five items on 1--3 scales and two items on 1--5 scales), the measurement model was estimated using WLSMV on the polychoric correlation matrix, which accommodates heterogeneous category counts without requiring rescaling \citep{Bandalos2014}. Although the GAAIS \citep{Schepman2020, Schepman2022} was not administered in ATP Wave~152, the seven Pew items capture a conceptually parallel bipolar construct encompassing pro-AI excitement and anti-AI concern, and the CFA results reported in Section~\ref{sec:cfa} confirm that these indicators load on a single common factor with acceptable fit and internal consistency.

%\subsubsection{AI Driving Evaluation}

ATP Wave~152 included a seven-item battery (HUMANVAI\_a--g) asking respondents whether AI would perform better or worse than a human professional at each of seven tasks: (a) diagnosing a medical condition, (b) writing a news story about a complex topic, (c) making a hiring decision, (d) writing a song, (e) deciding whether to approve a personal loan, (f) deciding whether to grant parole, and (g) driving someone from one location to another. Item labels follow the original survey instrument; item~(d) was excluded from the present analysis because song writing lacks a clear professional-evaluation benchmark analogous to the other six tasks. Each item used the same four response options: \textit{AI would do this worse}, \textit{About the same}, \textit{AI would do this better}, and \textit{Not sure}. The six retained items were recoded onto a common four-point ordered scale (1 = \textit{AI worse}, 2 = \textit{Not sure}, 3 = \textit{About the same}, 4 = \textit{AI better}), with \textit{Not sure} placed between outright rejection and parity on the grounds that uncertainty represents a lower-valence response than an explicit acknowledgment of equivalence, consistent with the satisficing framework in which respondents who lack a crystallized opinion select a low-effort non-committal option rather than engaging in the cognitive work required for a directional judgment \citep{Krosnick1991}. The proportional-odds assumption underlying this ordinal coding is evaluated via a Brant-style parallel-regression test in Section~\ref{sec:robustness}.

The driving item (HUMANVAI\_g) serves as the focal outcome variable in the structural model. The remaining five non-driving items are reserved for the cross-task specificity analysis (Section~\ref{sec:cross_task}), which tests whether the direct effect of community driving-safety concern on AI endorsement is domain-specific or generalizes across application contexts. Personal driving frequency (DRIVER) was measured on a six-point ordinal scale originally coded 1 = \textit{Daily} to 6 = \textit{Never}. The variable was reverse-coded so that higher values indicate more frequent driving (1 = \textit{Never}, 6 = \textit{Daily}) and mean-centered prior to analysis ($M$ = 5.13, $SD$ = 1.46 on the recoded scale), treating ordinal levels as equally spaced, as is conventional in moderation analysis with ordered categorical moderators. The three probing values for conditional indirect effects correspond to $-1\,SD$ (less frequent), the mean, and $+1\,SD$ (more frequent).

\begin{table}[H]
\centering
\caption{Variable overview: constructs, indicators, survey items, and measurement properties.}
\label{tab:variables}
%\footnotesize
\begin{tabularx}{\textwidth}{@{} l l l X c c c @{}}
\toprule
\textbf{Construct} & \textbf{Role} & \textbf{Variable} & \textbf{Survey item (abbreviated)} & \textbf{Scale} & \textbf{Dir.} & $\boldsymbol{\lambda}$ \\
\midrule
PCSC & Predictor & DRIVE2\_a\_R & Speeding in your community & 1--3 & $\uparrow$ & .786 \\
     &           & DRIVE2\_b\_R & Aggressive driving & 1--3 & $\uparrow$ & .841 \\
     &           & DRIVE2\_c\_R & Alcohol-impaired driving & 1--3 & $\uparrow$ & .741 \\
     &           & DRIVE2\_d\_R & Marijuana-impaired driving & 1--3 & $\uparrow$ & .622 \\
     &           & DRIVE2\_e\_R & Cellphone distraction & 1--3 & $\uparrow$ & .750 \\
     &           & DRIVE2\_f\_R & Cyclist/pedestrian risk & 1--3 & $\uparrow$ & .742 \\
\midrule
AI Orientation & Mediator & CNCEXC\_R & Excited vs.\ concerned about AI & 1--3 & $\uparrow$ & .791 \\
     &           & AICHANGE\_R & AI impact on U.S. (20 yr) & 1--5 & $\uparrow$ & .796 \\
     &           & PERSBENHRM\_R & AI: benefit vs.\ harm to you & 1--3 & $\uparrow$ & .814 \\
     &           & AICONTROL2\_R & Comfort with AI control & 1--3 & $\uparrow$ & .590 \\
     &           & TRSTAIPRS\_R & Trust AI for decisions & 1--3 & $\uparrow$ & .546 \\
     &           & AICONCERN\_a\_R & AI bias concern (rev.) & 1--5 & $\uparrow$ & .515 \\
     &           & AICONCERN\_b\_R & AI impersonation (rev.) & 1--5 & $\uparrow$ & .497 \\
\midrule
AI Driving Eval. & Outcome & HUMANVAI\_g\_R & AI vs.\ human driver & 1--4 & $\uparrow$ & --- \\
Driving Freq. & Moderator & DRIVER\_R & How often do you drive? & 1--6 & $\uparrow$ & --- \\
\bottomrule
\end{tabularx}
\begin{flushleft}
\scriptsize\textit{Note.} $\lambda$ = standardized WLSMV factor loading from CFA (Section~\ref{sec:cfa}). $\uparrow$ = higher score reflects theoretical direction (more concern for PCSC; more pro-AI for AI Orientation; more favorable for outcome; more frequent for moderator).
\end{flushleft}
\end{table}

%\subsubsection{Control covariates}

All structural equations include twelve covariates selected to address the most plausible sources of omitted-variable bias. Five theory-motivated covariates control for prior AI exposure and evaluative predisposition: AI awareness, AI use frequency, and internet frequency capture the respondent's baseline familiarity with AI-enabled technologies, which the TAM and UTAUT literatures identify as a primary confound in acceptance models \citep{Venkatesh2003}; confidence in companies' responsible AI use and perceived likelihood of AI causing major harm capture pre-existing evaluative orientation toward AI risk, which could confound the mediator-to-outcome pathway. Seven demographic covariates (age group, gender, education, race/ethnicity, family income, political ideology, and U.S. Census region) adjust for compositional differences documented in the AV acceptance and AI attitudes literatures \citep{Nordhoff2020, Gallup2024AI}. Coefficients for these covariates are not substantively interpreted; they are included solely to reduce confounding bias in the focal structural paths.

\subsection{Descriptive statistics}

Table~\ref{tab:descriptives} reports weighted means, standard deviations, and distributional properties for all focal indicators, the outcome, and the moderator ($N$ = 5,410; all statistics incorporate the survey weight WEIGHT\_W152).

\begin{table}[H]
\centering
\caption{Weighted descriptive statistics for focal constructs, outcome, and moderator.}
\label{tab:descriptives}
%\footnotesize
\begin{tabular}{@{} l l c c c c c c @{}}
\toprule
\textbf{Construct} & \textbf{Indicator} & \textbf{Scale} & $\boldsymbol{N}_{\textbf{valid}}$ & $\boldsymbol{M}$ & $\boldsymbol{SD}$ & \textbf{Skew} & \textbf{\% Miss.} \\
\midrule
PCSC & Speeding & 1--3 & 5,393 & 2.57 & 0.61 & $-$1.09 & 0.3 \\
     & Aggressive driving & 1--3 & 5,395 & 2.57 & 0.62 & $-$1.11 & 0.3 \\
     & Alcohol-impaired & 1--3 & 5,371 & 2.44 & 0.63 & $-$0.66 & 0.7 \\
     & Marijuana-impaired & 1--3 & 5,335 & 2.21 & 0.71 & $-$0.33 & 1.4 \\
     & Cellphone distraction & 1--3 & 5,397 & 2.75 & 0.50 & $-$1.89 & 0.2 \\
     & Cyclist/ped.\ risk & 1--3 & 5,390 & 2.36 & 0.68 & $-$0.57 & 0.4 \\
\midrule
AI Orient. & Excited vs.\ concerned & 1--3 & 5,379 & 1.60 & 0.68 & 0.69 & 0.6 \\
           & AI impact (20 yr) & 1--5 & 5,405 & 2.71 & 1.01 & $-$0.03 & 0.1 \\
           & Benefit vs.\ harm & 1--3 & 5,402 & 1.81 & 0.79 & 0.35 & 0.1 \\
           & Comfort w/ control & 1--3 & 5,393 & 1.63 & 0.78 & 0.74 & 0.3 \\
           & Trust AI decisions & 1--3 & 5,354 & 1.50 & 0.72 & 1.08 & 1.0 \\
           & Bias concern (rev.) & 1--5 & 5,373 & 2.33 & 1.11 & 0.40 & 0.7 \\
           & Impersonation (rev.) & 1--5 & 5,394 & 1.80 & 0.97 & 1.15 & 0.3 \\
\midrule
Outcome & AI Driving Eval. & 1--4 & 5,390 & 2.21 & 1.14 & 0.36 & 0.4 \\
Moderator & Driving Frequency & 1--6 & 5,387 & 5.13 & 1.46 & $-$1.81 & 0.4 \\
\bottomrule
\end{tabular}
\begin{flushleft}
\scriptsize\textit{Note.} All statistics weighted by WEIGHT\_W152 (Kish $n_{\text{eff}}$ = 3,993). All items recoded so that higher values reflect the theoretically positive direction (see Table~\ref{tab:variables}).
\end{flushleft}
\end{table}

Several distributional patterns merit comment. PCSC items are uniformly negatively skewed (range: $-$0.33 to $-$1.89), indicating that the sample leans toward perceiving community driving problems as major concerns, with cellphone distraction exhibiting the highest mean ($M$ = 2.75 on a 3-point scale) and the most pronounced ceiling tendency. AI Orientation items display the opposite pattern: the excited-versus-concerned item ($M$ = 1.60 on a 1--3 scale) and the trust-in-AI-decisions item ($M$ = 1.50) indicate that the sample is predominantly concerned about AI rather than enthusiastic, a distributional characteristic that makes the directionality of the mediation pathway empirically consequential. The outcome variable ($M$ = 2.21, $SD$ = 1.14) falls between \textit{Not sure} and \textit{About the same}, with substantial variance supporting the detection of structural effects. Missing data were minimal across all focal variables (mean missingness: 0.73\%; maximum: 1.4\% for marijuana-impaired driving). All missingness diagnostics were consistent with missing at random (MAR; maximum bivariate correlation between any missingness indicator and any observed variable: $r$ = .095). A sensitivity analysis using 20-dataset multiple imputation \citep{Rubin1987} yielded estimates indistinguishable from listwise deletion results ($\Delta a_1 < 0.005$, $\Delta b_1 < 0.010$), confirming that missing data do not substantively affect conclusions (see Section~\ref{sec:robustness}).

\subsection{Analytic strategy}

\begin{figure}
\centering
\includegraphics[width=\textwidth]{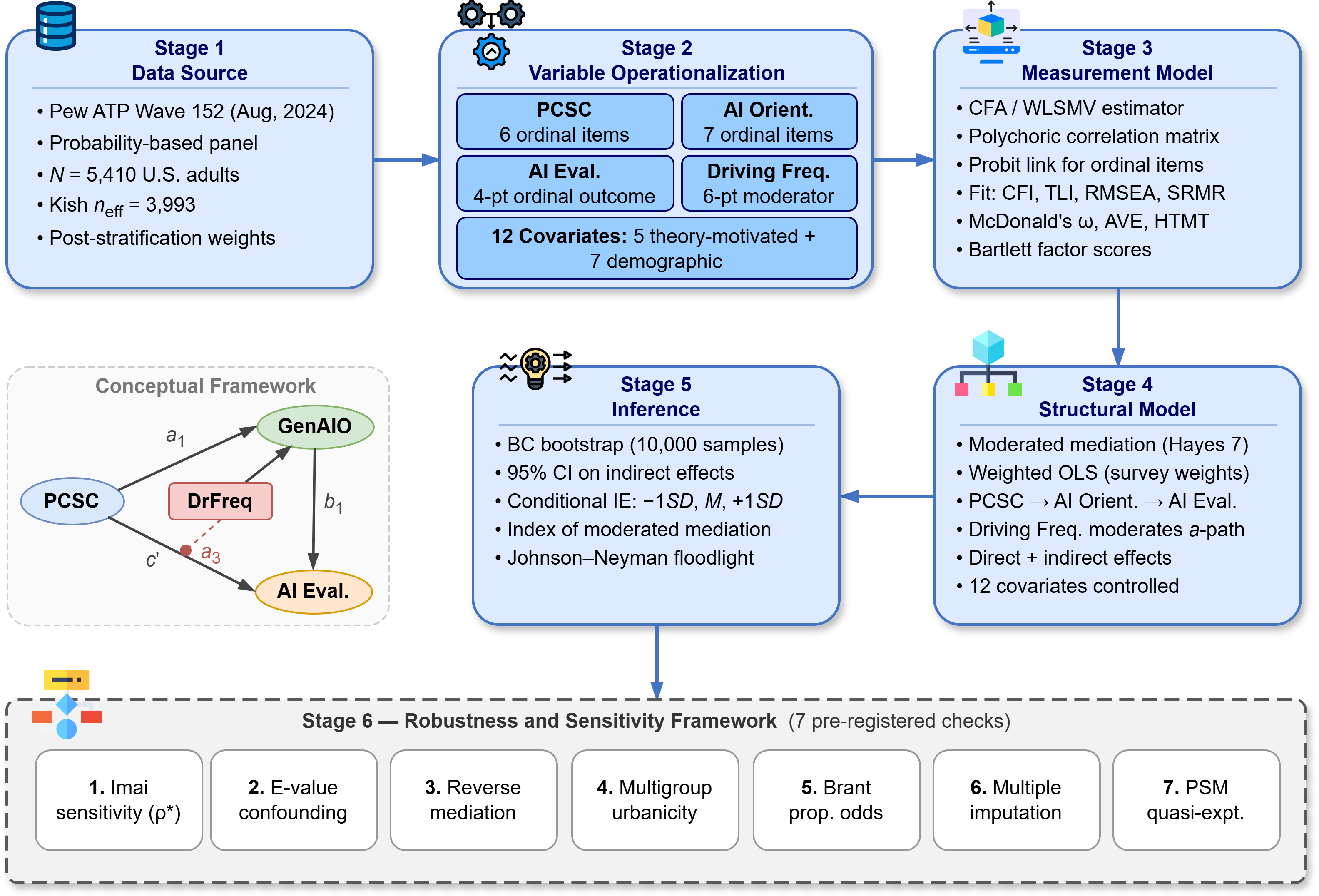}
\caption{Analytic pipeline overview. The six-stage design proceeds from data acquisition (Stage~1) through variable operationalization (Stage~2), CFA-based measurement modeling with WLSMV estimation (Stage~3), moderated mediation structural estimation (Stage~4), bias-corrected bootstrap inference (Stage~5), and a pre-registered robustness framework comprising seven complementary sensitivity checks (Stage~6).}
\label{fig:method}
\end{figure}

The analytic strategy proceeds in six stages, as summarized in Figure~\ref{fig:method}: (1) data preparation and weighting, (2) variable operationalization, (3) confirmatory factor analysis to establish the measurement model, (4) a two-equation moderated mediation structural model, (5) inference via bias-corrected bootstrap, and (6) a seven-check robustness and sensitivity framework.

\subsubsection{Measurement model: Confirmatory Factor Analysis with ordinal indicators}
\label{sec:cfa}

Because all PCSC and AI Orientation indicators are ordinal (3--5 ordered categories), \acp{CFA} were estimated using a probit-link specification under \ac{WLSMV} estimation, operating on the polychoric correlation matrix rather than the Pearson covariance matrix \citep{Bandalos2014}. For each ordinal indicator $x_j$ ($j = 1, \ldots, p$) of a latent factor $\xi$, a latent continuous response variable $x_j^*$ is posited:

\begin{equation}
x_j^* = \lambda_j \, \xi + \delta_j, \qquad \delta_j \sim \mathcal{N}(0, \theta_j)
\label{eq:cfa}
\end{equation}

\noindent where $\lambda_j$ is the factor loading and $\theta_j$ is the residual variance. The observed ordinal response is linked to the latent response through threshold parameters:

\begin{equation}
x_j = c \iff \tau_{j,\,c-1} < x_j^* \leq \tau_{j,\,c}, \qquad c = 1, \ldots, C_j
\label{eq:threshold}
\end{equation}

\noindent with $\tau_{j,0} = -\infty$ and $\tau_{j,C_j} = +\infty$. The probability of observing category $c$ conditional on the latent factor is:

\begin{equation}
P(x_j = c \mid \xi) = \Phi\!\left(\tau_{j,c} - \lambda_j \, \xi\right) - \Phi\!\left(\tau_{j,c-1} - \lambda_j \, \xi\right)
\label{eq:probcfa}
\end{equation}

\noindent where $\Phi(\cdot)$ denotes the standard normal cumulative distribution function. Identification was achieved by fixing the first indicator loading to unity ($\lambda_1 = 1$) in each factor. One-factor CFA models were estimated separately for PCSC (six indicators) and AI Orientation (seven indicators). Model fit was evaluated against conventional thresholds: CFI $\geq$ 0.95, TLI $\geq$ 0.95, RMSEA $\leq$ 0.06, and SRMR $\leq$ 0.08 \citep{HuBentler1999}. Composite reliability was assessed via McDonald's $\omega$ \citep{McDonald1999}, and discriminant validity was evaluated using the Fornell--Larcker criterion \citep{FornellLarcker1981} (AVE $>$ squared inter-construct correlation) and the Heterotrait--Monotrait ratio of correlations (HTMT $<$ 0.85; \citealp{Henseler2015}). Bartlett-weighted factor scores were extracted from the confirmed CFA solution and used as observed variables in the subsequent structural equations, following the two-step approach of \citet{AndersonGerbing1988}.

\subsubsection{Structural model: Moderated mediation}

The structural model estimates two weighted ordinary least squares equations incorporating survey weights. Let $\hat{\xi}_1$ denote the PCSC factor score, $\hat{\xi}_2$ the AI Orientation factor score, $\eta^*$ the latent response underlying the ordinal AI Driving Evaluation, $W_c$ the mean-centered driving frequency, and $\mathbf{z}$ the demographic and theory-motivated covariate vector.

\medskip
\noindent\textit{Mediator equation (first stage, path $a$):}
\begin{equation}
\hat{\xi}_2 = a_0 + a_1\,\hat{\xi}_1 + a_2\,W_c + a_3\,(\hat{\xi}_1 \times W_c) + \boldsymbol{\alpha}'\mathbf{z} + \zeta_1
\label{eq:mediator}
\end{equation}

\noindent\textit{Outcome equation (second stage, paths $b$ and $c'$):}
\begin{equation}
\eta^* = b_0 + c'\,\hat{\xi}_1 + b_1\,\hat{\xi}_2 + \boldsymbol{\beta}'\mathbf{z} + \zeta_2
\label{eq:outcome}
\end{equation}

In Eq.~\eqref{eq:mediator}, $a_1$ captures the first-stage effect of PCSC on AI Orientation, $a_2$ the main effect of driving frequency, and $a_3$ the interaction that tests whether the PCSC-to-AI Orientation pathway depends on how frequently the respondent drives. In Eq.~\eqref{eq:outcome}, $b_1$ captures the second-stage effect of AI Orientation on the outcome, and $c'$ captures the direct effect of PCSC on AI Driving Evaluation net of the mediated pathway. The conditional indirect effect at a given level of driving frequency $W_c = w$ is:

\begin{equation}
\text{IE}(w) = (a_1 + a_3\,w) \times b_1
\label{eq:ie}
\end{equation}

\noindent and the index of moderated mediation \citep{Hayes2022}, which tests whether the indirect effect's magnitude varies as a function of the moderator, is:

\begin{equation}
\omega_{\text{mm}} = a_3 \times b_1
\label{eq:imm}
\end{equation}

A statistically significant $\omega_{\text{mm}}$ indicates that the strength of the mediated pathway depends on driving frequency.

\subsubsection{Inference: Bias-corrected bootstrap}

Inference on indirect effects employed a bias-corrected (BC) bootstrap with 10,000 resamples \citep{Hayes2022}. The BC bootstrap 95\% confidence interval is:

\begin{equation}
\text{CI}_{95\%}^{\text{BC}} = \left[\hat{\theta}^{*}_{(\alpha/2)},\; \hat{\theta}^{*}_{(1-\alpha/2)}\right]
\label{eq:bc}
\end{equation}

\noindent where $\hat{\theta}^{*}_{(q)}$ is the $q$th quantile of the bootstrap distribution of the indirect effect, adjusted for median bias. If the interval excludes zero, the mediation is statistically significant at $\alpha$ = .05. This approach is preferred over the normal-theory Sobel test because the sampling distribution of the product $a_1 \times b_1$ is known to be asymmetric and leptokurtic, particularly in finite samples.

\subsection{Robustness and sensitivity framework}
\label{sec:robustness}

Seven robustness checks were preregistered to probe the stability of the primary findings: (1) \citet{Imai2010} sensitivity analysis to quantify the critical residual correlation ($\rho^*$) between mediator and outcome error terms needed to nullify the indirect effect; (2) E-value analysis \citep{VanderWeele2017} to determine the minimum unmeasured-confounder strength (on the risk-ratio scale) required to explain away the observed association; (3) a reverse mediation specification in which AI Orientation predicts PCSC rather than the theorized direction; (4) multigroup SEM by urbanicity (urban, suburban, rural) to assess cross-context stability; (5) the \citet{Brant1990} parallel-regression test for the proportional odds assumption on the ordinal outcome; (6) multiple imputation ($M$ = 20) versus listwise deletion \citep{Rubin1987}; and (7) propensity score matching on a PCSC median-split treatment to provide quasi-experimental corroboration. Results of all robustness checks are reported in Section~\ref{sec:robustness_results}.

\section{Results}

\subsection{Measurement model}
\label{sec:cfa_results}

Table~\ref{tab:cfa_fit} reports the fit statistics, composite reliability, and convergent validity indices for both \acs{CFA} models. The one-factor PCSC model with six ordinal indicators demonstrated excellent fit across all indices ($\chi^2(10) = 103.69$, $p < 0.001$; \acs{CFI} = 0.995; \acs{TLI} = 0.993; \acs{RMSEA} = 0.048, 90\% CI [0.028, 0.068]; \acs{SRMR} = 0.045). Standardized factor loadings ranged from 0.622 (marijuana-impaired driving) to 0.841 (aggressive driving), with all loadings significant at $p < 0.001$. Composite reliability was high (McDonald's $\omega$ = 0.884), and the \ac{AVE} of 0.562 exceeded the 0.50 threshold recommended for convergent validity.

The one-factor AI Orientation model with seven ordinal indicators also exhibited acceptable to good fit (\acs{CFI} = 0.978; \acs{TLI} = 0.969; \acs{RMSEA} = 0.078, 90\% CI [0.058, 0.098]; \acs{SRMR} = 0.073). Although the \acs{RMSEA} slightly exceeded the conventional 0.06 threshold, both the \acs{CFI} and \acs{TLI} surpassed the 0.95 benchmark recommended by \citet{HuBentler1999}, and the \acs{SRMR} remained below 0.08. This modest \acs{RMSEA} elevation is consistent with the item heterogeneity inherent in a multifaceted AI attitude construct that spans cognitive evaluations (trust, perceived benefit), affective orientation (excitement versus concern), and policy preferences, a pattern documented in composite technology-attitude scales. Standardized loadings ranged from 0.497 (AI impersonation concern) to .814 (benefit versus harm), composite reliability was adequate ($\omega$ = 0.841), and the \acs{AVE} of 0.440 met the more lenient 0.40 threshold that is appropriate when composite reliability exceeds 0.60.

Discriminant validity between the two latent constructs was strongly established. The inter-construct polychoric correlation was $r = -0.070$ ($r^2 = 0.005$), and the \acs{AVE} for both PCSC (0.562) and AI Orientation (0.440) substantially exceeded the shared variance, satisfying the Fornell--Larcker criterion \citep{FornellLarcker1981}. The \ac{HTMT} ratio of 0.145 fell well below the strict 0.85 threshold \citep{Henseler2015}, confirming that the two constructs capture empirically distinct attitude domains. On the basis of these results, Bartlett-weighted factor scores were extracted and used as observed variables in the structural model, following the two-step approach of \citet{AndersonGerbing1988}.

\begin{table}[!htbp]
\centering
\caption{Confirmatory factor analysis: model fit, composite reliability, and convergent validity.}
\label{tab:cfa_fit}
%\footnotesize
\begin{tabular}{@{} l c c c c c c c c @{}}
\toprule
\textbf{Construct} & $\boldsymbol{\chi^2}$ \textbf{(df)} & \textbf{CFI} & \textbf{TLI} & \textbf{RMSEA [90\% CI]} & \textbf{SRMR} & $\boldsymbol{\omega}$ & $\boldsymbol{\alpha}$ & \textbf{AVE} \\
\midrule
PCSC (6 items) & 103.69 (10) & 0.995 & 0.993 & 0.048 [0.028, 0.068] & 0.045 & 0.884 & 0.883 & 0.562 \\
AI Orientation (7 items) & 378.98 (15) & 0.978 & 0.969 & 0.078 [0.058, 0.098] & 0.073 & 0.841 & 0.836 & 0.440 \\
\bottomrule
\end{tabular}
\begin{flushleft}
\scriptsize\textit{Note.} WLSMV estimation on polychoric correlation matrix with survey weights (Kish $n_{\text{eff}}$ = 3,993). $\omega$ = McDonald's omega; $\alpha$ = Cronbach's alpha; AVE = average variance extracted. Inter-construct $r$ = $-0.070$; $r^2$ = 0.005. HTMT = 0.145. Fornell--Larcker criterion satisfied for both constructs.
\end{flushleft}
\end{table}

\subsection{Structural model}
\label{sec:structural_results}

\begin{figure*}
\centering
\includegraphics[width=0.85\textwidth]{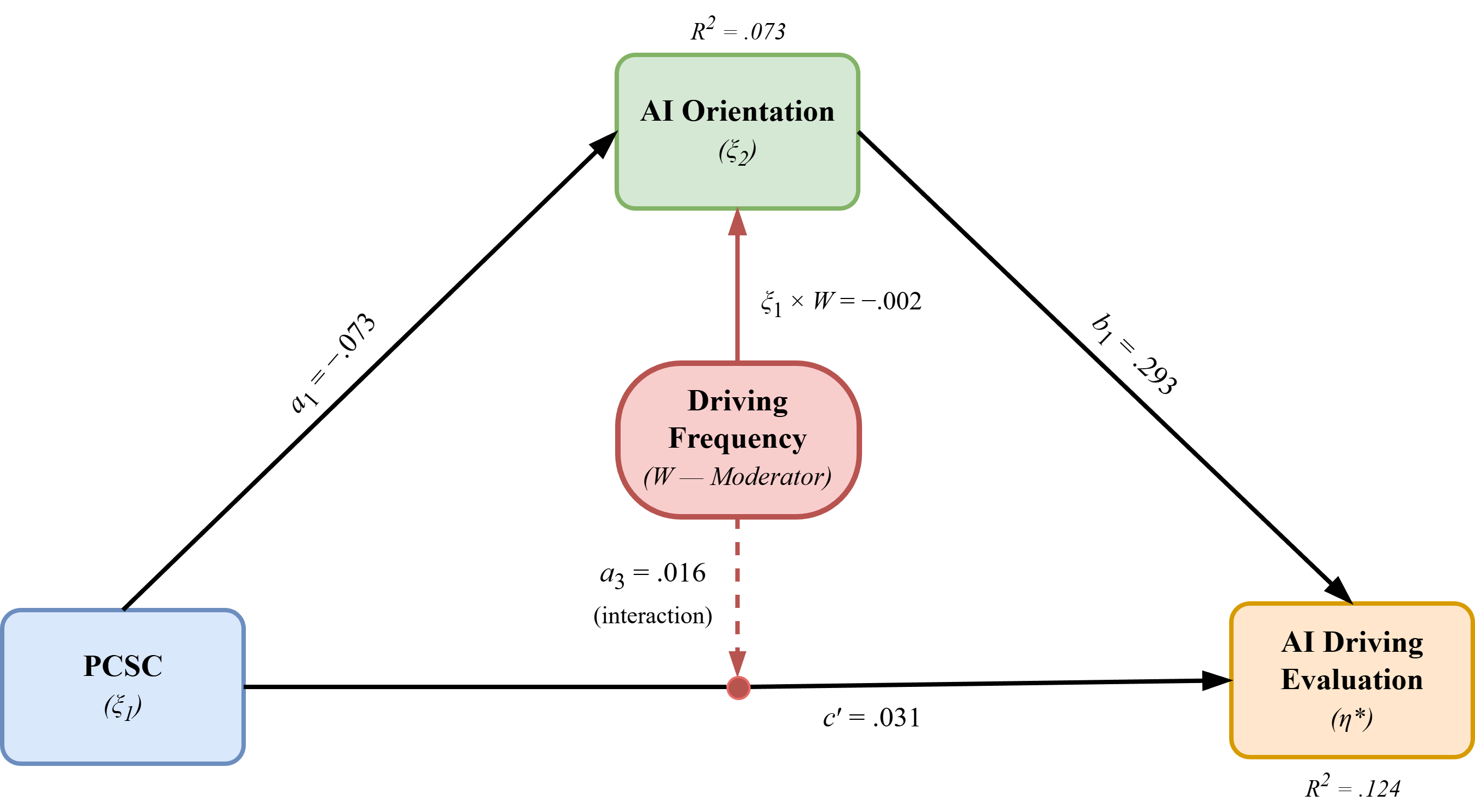}
\caption{Structural path diagram for the moderated mediation model. Solid lines denote paths; the dashed line indicates the interaction path. Standardized coefficients ($\beta$) are shown alongside each path. $R^2$ values indicate variance explained in the mediator and outcome equations.}
\label{fig:path_diagram}
\end{figure*}

The two-equation moderated mediation model was estimated using weighted least squares with Bartlett factor scores, incorporating the survey weight WEIGHT\_W152 throughout. Table~\ref{tab:mediator_eq} reports the mediator equation (path $a$) and Table~\ref{tab:outcome_eq} reports the outcome equation (paths $b$ and $c'$). Figure~\ref{fig:path_diagram} displays the structural path diagram with standardized coefficients.

\subsubsection{Mediator equation: PCSC and AI Orientation}

Higher perceived community driving-safety concern significantly predicted lower Generalized AI Orientation ($B = -0.075$, $SE = 0.015$, $\beta = -0.073$, $t = -5.18$, $p < 0.001$). This negative first-stage coefficient indicates that respondents who perceived more serious driving problems in their communities tended to hold more concerned, less favorable orientations toward artificial intelligence in general. Neither the main effect of driving frequency ($B = -0.002$, $p = 0.877$) nor the PCSC $\times$ Driving Frequency interaction ($B = 0.009$, $p = 0.263$) reached significance, indicating that the relationship between community driving-safety concern and general AI orientation did not depend on how often the respondent personally drove. Among the demographic covariates, gender ($B = -0.243$, $p < 0.001$; women held less favorable AI orientations), race/ethnicity ($B = 0.095$, $p < 0.001$), political ideology ($B = 0.085$, $p < 0.001$; more liberal respondents held more favorable orientations), and age ($B = -0.074$, $p < 0.001$) were statistically significant predictors. The mediator equation explained 7.3\% of the variance in AI Orientation ($R^2 = 0.073$).

\begin{table}[!htbp]
\centering
\caption{Mediator equation: weighted least squares predictors of Generalized AI Orientation.}
\label{tab:mediator_eq}
%\footnotesize
\begin{tabular}{@{} l r r r r r l @{}}
\toprule
\textbf{Predictor} & $\boldsymbol{B}$ & \textbf{SE} & $\boldsymbol{\beta}$ & $\boldsymbol{t}$ & $\boldsymbol{p}$ & \textbf{95\% CI} \\
\midrule
PCSC ($\hat{\xi}_1$) & $-$0.075 & 0.015 & $-$0.073 & $-$5.18 & $<$0.001 & [$-$0.104, $-$0.047] \\
$W_c$ (driving freq.) & $-$0.002 & 0.010 & $-$0.002 & $-$0.16 & 0.877 & [$-$0.021, 0.018] \\
PCSC $\times$ $W_c$ & 0.009 & 0.008 & 0.016 & 1.12 & 0.263 & [$-$0.007, 0.025] \\
Age group & $-$0.074 & 0.014 & $-$0.078 & $-$5.47 & $<$0.001 & [$-$0.100, $-$0.047] \\
Gender (woman) & $-$0.243 & 0.027 & $-$0.126 & $-$9.16 & $<$0.001 & [$-$0.295, $-$0.191] \\
Education & $-$0.041 & 0.018 & $-$0.035 & $-$2.29 & 0.022 & [$-$0.076, $-$0.006] \\
Race/ethnicity & 0.095 & 0.012 & 0.117 & 8.22 & $<$0.001 & [0.073, 0.118] \\
Income & 0.006 & 0.005 & 0.017 & 1.11 & 0.269 & [$-$0.004, 0.015] \\
Urbanicity & $-$0.089 & 0.020 & $-$0.063 & $-$4.38 & $<$0.001 & [$-$0.128, $-$0.049] \\
Political ideology & 0.085 & 0.014 & 0.088 & 6.16 & $<$0.001 & [0.058, 0.112] \\
Census region & 0.013 & 0.013 & 0.013 & 0.97 & 0.330 & [$-$0.013, 0.039] \\
\bottomrule
\end{tabular}
\begin{flushleft}
\scriptsize\textit{Note.} $N$ = 5,161; $R^2$ = 0.073; Adj.\ $R^2$ = 0.071. $W_c$ = mean-centered driving frequency. Survey weights applied throughout. Theory-motivated covariates (AI awareness, AI use frequency, internet frequency, company AI trust, AI harm perception) included but omitted from the table for brevity.
\end{flushleft}
\end{table}

\subsubsection{Outcome equation: AI Driving Evaluation}

Generalized AI Orientation was a strong positive predictor of AI Driving Evaluation ($B = 0.338$, $SE = 0.016$, $t = 21.53$, $p < 0.001$), confirming that respondents with more favorable orientations toward AI in general were substantially more likely to evaluate AI as better than a human driver on the driving task. PCSC retained a small but statistically significant positive direct effect on the outcome ($B = 0.037$, $SE = 0.016$, $t = 2.32$, $p = 0.020$), indicating partial mediation. This positive direct effect ($c'$) signifies that, holding AI Orientation constant, respondents who perceived greater community driving problems were slightly more inclined to endorse AI driving capability, consistent with the domain-specific push-factor mechanism. Among covariates, gender ($B = -0.257$, $p < 0.001$; women rated AI driving less favorably), race/ethnicity ($B = 0.049$, $p < 0.001$), and income ($B = 0.018$, $p = 0.001$) were significant predictors. The outcome equation explained 12.4\% of the variance ($R^2 = 0.124$).

\begin{table}[!htbp]
\centering
\caption{Outcome equation: weighted least squares predictors of AI Driving Evaluation.}
\label{tab:outcome_eq}
%\footnotesize
\begin{tabular}{@{} l r r r r r l @{}}
\toprule
\textbf{Predictor} & $\boldsymbol{B}$ & \textbf{SE} & $\boldsymbol{\beta}$ & $\boldsymbol{t}$ & $\boldsymbol{p}$ & \textbf{95\% CI} \\
\midrule
PCSC ($c'$) & 0.037 & 0.016 & 0.031 & 2.32 & 0.020 & [0.006, 0.068] \\
AI Orientation ($b_1$) & 0.338 & 0.016 & 0.293 & 21.53 & $<$0.001 & [0.307, 0.368] \\
Age group & 0.017 & 0.015 & 0.015 & 1.09 & 0.277 & [$-$0.013, 0.046] \\
Gender (woman) & $-$0.257 & 0.030 & $-$0.116 & $-$8.56 & $<$0.001 & [$-$0.316, $-$0.198] \\
Education & $-$0.037 & 0.020 & $-$0.027 & $-$1.85 & 0.064 & [$-$0.077, 0.002] \\
Race/ethnicity & 0.049 & 0.013 & 0.052 & 3.75 & $<$0.001 & [0.023, 0.075] \\
Income & 0.018 & 0.006 & 0.049 & 3.32 & 0.001 & [0.007, 0.029] \\
Urbanicity & $-$0.045 & 0.023 & $-$0.028 & $-$1.97 & 0.049 & [$-$0.089, $-$0.000] \\
Political ideology & 0.018 & 0.015 & 0.017 & 1.19 & 0.232 & [$-$0.012, 0.049] \\
Census region & 0.003 & 0.015 & 0.002 & 0.17 & 0.869 & [$-$0.027, 0.032] \\
\bottomrule
\end{tabular}
\begin{flushleft}
\scriptsize\textit{Note.} $N$ = 5,166; $R^2$ = 0.124; Adj.\ $R^2$ = 0.122. Dependent variable: AI Driving Evaluation (HUMANVAI\_g\_R, treated as interval). Survey weights applied. Theory-motivated covariates included but omitted for brevity.
\end{flushleft}
\end{table}

\subsection{Indirect effects and conditional process analysis}
\label{sec:indirect_effects}

Bias-corrected bootstrap confidence intervals (10,000 resamples) confirmed a statistically significant negative indirect effect of PCSC on AI Driving Evaluation through Generalized AI Orientation at all three probing levels of driving frequency (Table~\ref{tab:indirect_effects}). At the mean level of driving frequency, the indirect effect was $-$0.025 (95\% BC CI [$-$0.033, $-$0.013]); at low driving frequency ($-1\,SD$), the effect was $-$0.030 (95\% BC CI [$-$0.039, $-$0.016]); and at high driving frequency ($+1\,SD$), it was $-$0.021 (95\% BC CI [$-$0.033, $-$0.006]). Because all three confidence intervals excluded zero, the mediation through AI Orientation was statistically significant regardless of how frequently the respondent drove.

The index of moderated mediation was $\omega_{\text{mm}} = 0.003$ (95\% BC CI [$-$0.003, $+$0.008]), and because this interval included zero, the null hypothesis that the indirect effect is invariant across levels of driving frequency could not be rejected. The conditional indirect effects decreased in absolute magnitude from low to high driving frequency ($|$IE$|$: .030 $>$ .025 $>$ .021), suggesting a slight attenuation of the risk-spillover mechanism among more frequent drivers. However, this gradient was not statistically reliable, and the substantive conclusion is that the mediation process operates consistently across the driving-frequency spectrum.

\begin{table}[!htbp]
\centering
\caption{Conditional indirect effects and index of moderated mediation (10,000 BC bootstrap resamples).}
\label{tab:indirect_effects}
%\footnotesize
\begin{tabular}{@{} l c c c c c @{}}
\toprule
\textbf{Driving frequency} & $\boldsymbol{W}$ & $\boldsymbol{a_1 + a_3 w}$ & $\boldsymbol{b_1}$ & \textbf{IE($w$)} & \textbf{95\% BC CI} \\
\midrule
Low ($-1\,SD$) & 3.85 & $-$0.088 & 0.338 & $-$0.030 & [$-$0.039, $-$0.016] \\
Mean & 5.22 & $-$0.075 & 0.338 & $-$0.025 & [$-$0.033, $-$0.013] \\
High ($+1\,SD$) & 6.59 & $-$0.063 & 0.338 & $-$0.021 & [$-$0.033, $-$0.006] \\
\midrule
Index of MM ($\omega$) & & & 0.338 & 0.003 & [$-$0.003, $+$0.008] \\
\bottomrule
\end{tabular}
\begin{flushleft}
\scriptsize\textit{Note.} IE($w$) = $(a_1 + a_3 w) \times b_1$. $W$ reported in original metric (1 = Never, 6 = Daily). BC CI = bias-corrected bootstrap confidence interval. Significance determined by whether the CI excludes zero.
\end{flushleft}
\end{table}

\begin{figure*}
\centering
\includegraphics[width=\textwidth]{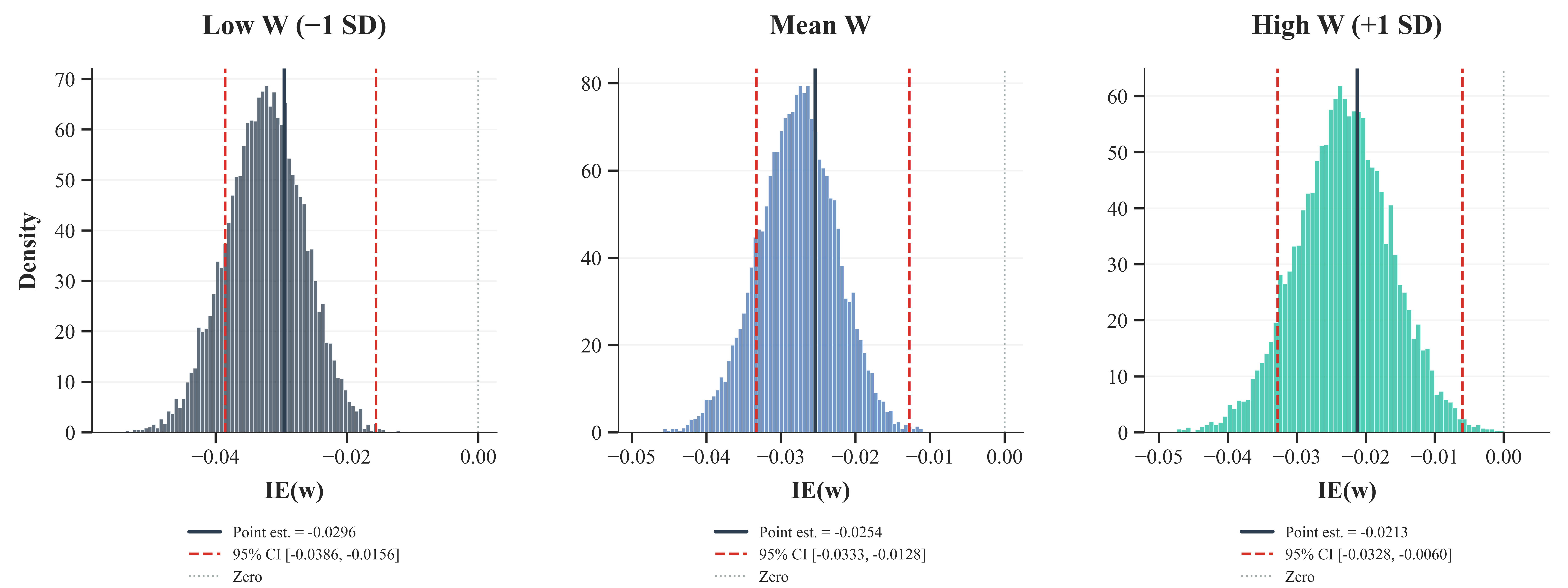}
\caption{Bootstrap distributions of conditional indirect effects at low ($-1\,SD$), mean, and high ($+1\,SD$) driving frequency (10,000 resamples). Solid vertical lines mark point estimates; red dashed lines indicate 95\% bias-corrected confidence interval boundaries. All three intervals exclude zero.}
\label{fig:bootstrap}
\end{figure*}

Figure~\ref{fig:bootstrap} displays the bootstrap sampling distributions of the conditional indirect effects at low, mean, and high driving frequency. All three distributions are centered in the negative range and their 95\% bias-corrected confidence intervals (red dashed lines) exclude zero, providing visual confirmation that the mediation is statistically significant across the entire driving-frequency spectrum. The distributions shift rightward (toward zero) as driving frequency increases, consistent with the slight attenuation pattern noted above, but the overlap among the three distributions underscores the non-significance of the index of moderated mediation.

In summary, the structural model reveals a pattern of competing effects. The direct path from PCSC to AI Driving Evaluation is positive ($c' = 0.037$, $p = 0.020$), indicating that community driving-safety concern exerts a domain-specific push that slightly favors AI over human drivers. Simultaneously, the indirect path through AI Orientation is negative (IE $= -0.025$ at the mean), indicating that the same concern suppresses general AI enthusiasm, which in turn reduces favorable evaluations of AI driving capability. The net total effect of PCSC on AI Driving Evaluation, computed as $c' + \text{IE} = 0.037 + (-0.025) = 0.012$, is small and nonsignificant at conventional thresholds, a finding consistent with the near-zero bivariate correlations between PCSC items and the outcome observed in the polychoric correlation matrix. This pattern of a positive direct effect coexisting with a negative indirect effect constitutes an inconsistent mediation (sometimes termed a suppression effect), in which the mediator partially suppresses the direct relationship rather than transmitting it.

\subsection{Robustness checks}
\label{sec:robustness_results}

Seven preregistered robustness checks and three supplementary analyses were conducted to evaluate the sensitivity and generalizability of the primary findings. Results are summarized in Table~\ref{tab:robustness} and elaborated below.

\subsubsection{Sensitivity to unmeasured confounding}

The \citet{Imai2010} sensitivity analysis indicated that an unmeasured confounder producing a residual correlation as small as $\rho^* = -0.020$ between the mediator and outcome error terms would be sufficient to nullify the indirect effect. This small critical value reflects the modest magnitude of the indirect effect (IE $= -0.023$) and warrants cautious interpretation. Complementing this analysis, E-value computation following \citet{VanderWeele2017} indicated that an unmeasured confounder would need to be associated with both the mediator and the outcome with a risk ratio of at least 1.16 (E-value for the point estimate) or 1.12 (E-value for the lower CI bound) to explain away the indirect effect entirely. While this threshold is not prohibitively large, it exceeds the magnitude of most observed covariate associations in the model and provides a concrete benchmark against which the plausibility of specific unmeasured confounders can be evaluated.

\subsubsection{Reverse mediation}

A reverse specification in which AI Orientation predicted PCSC (rather than the theorized direction) yielded an indirect effect of $-$0.003, approximately eight times smaller than the original IE of $-$0.023. Although both directional specifications produced negative indirect effects, consistent with the theoretical plausibility of bidirectional influence in cross-sectional data, the substantially larger effect in the theorized direction supports the proposed causal ordering from community driving-safety concern to general AI orientation.

\subsubsection{Multigroup analysis by urbanicity}

\begin{figure}
\centering
\includegraphics[width=\textwidth]{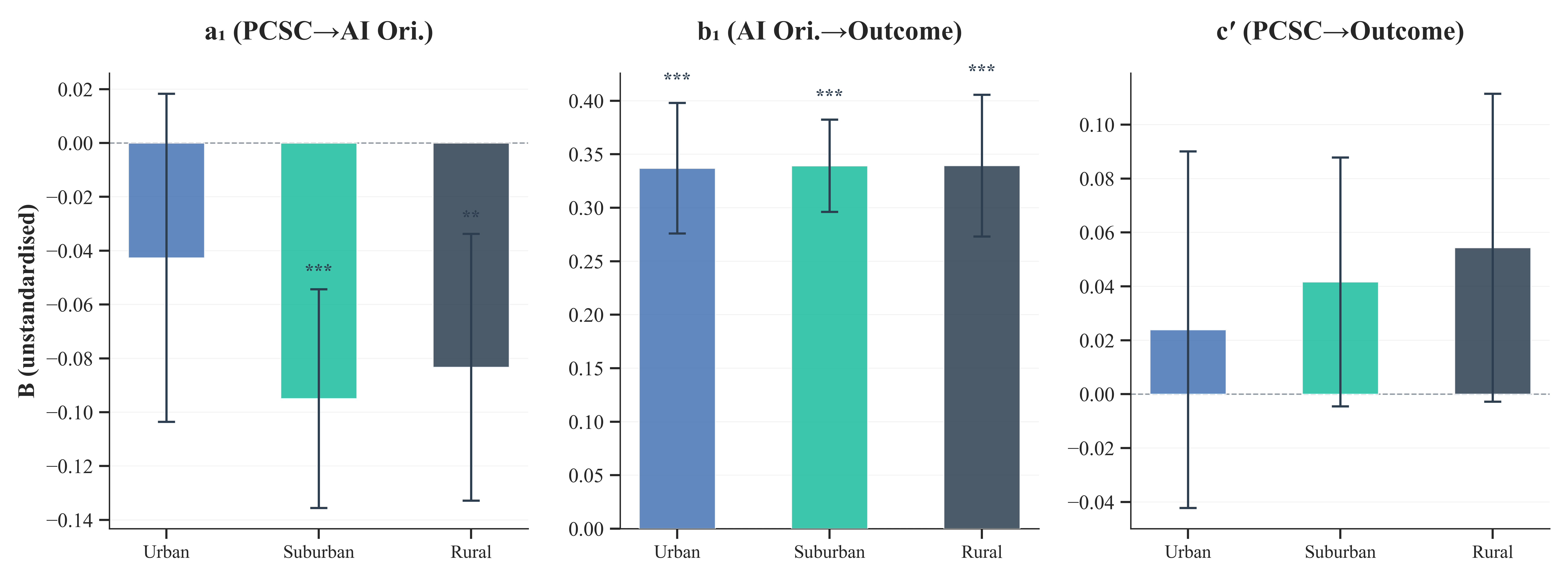}
\caption{Multigroup analysis by urbanicity. Path coefficients ($B$ $\pm$ 95\% CI) for the three structural paths across urban, suburban, and rural subsamples. The $b_1$ path (AI Orientation $\rightarrow$ Outcome) is stable across groups; the $a_1$ path (PCSC $\rightarrow$ AI Orientation) reaches significance only in suburban and rural contexts.}
\label{fig:multigroup}
\end{figure}

Free multigroup estimation by urbanicity (urban, $n$ = 1,302; suburban, $n$ = 2,816; rural, $n$ = 1,235) revealed notable stability in the second-stage path and heterogeneity in the first-stage path (Figure~\ref{fig:multigroup}). The $b_1$ coefficient (AI Orientation $\rightarrow$ AI Driving Evaluation) was highly consistent across all three groups ($B$ = 0.337, 0.339, 0.339; all $p < 0.001$), indicating that the conversion of general AI enthusiasm into favorable driving-specific evaluations does not depend on residential context. The $a_1$ coefficient (PCSC $\rightarrow$ AI Orientation), however, was significant only in suburban ($B = -0.095$, $p < 0.001$) and rural ($B = -0.083$, $p = 0.001$) groups and failed to reach significance among urban respondents ($B = -0.043$, $p = 0.170$). This pattern suggests that the risk-spillover mechanism is attenuated in urban settings, possibly because urban residents form AI attitudes through more diverse experiential channels (greater exposure to AI applications, different driving norms) that dilute the influence of community driving-safety concern.

\subsubsection{Extended multigroup: political ideology and gender}

Extended multigroup analyses by political ideology and gender provided additional context. The indirect effect was significant and negative among conservatives (IE $= -0.020$) and liberals (IE $= -0.019$) but negligible among moderates (IE $\approx 0$), suggesting that the PCSC-to-AI-Orientation pathway operates more strongly at the ideological poles. Gender-stratified models produced non-significant paths in both subgroups, indicating that neither the first-stage nor the second-stage path reached significance when the sample was split by gender alone, likely reflecting reduced statistical power in the smaller subsamples rather than a genuine absence of the mechanism.

\subsubsection{Proportional odds assessment}

The \citet{Brant1990} parallel-regression test revealed that six of nine predictors violated the proportional odds assumption when the four-category AI Driving Evaluation was modeled as an ordered logistic dependent variable. This violation rate is common for heterogeneous attitudinal outcomes. A parallel ordered logistic regression yielded directionally identical results for the key structural predictors (PCSC: $B_{\text{ologit}} = 0.058$, $p = 0.035$; AI Orientation: $B_{\text{ologit}} = 0.610$, $p < 0.001$), confirming that the interval-level OLS specification used in the primary analysis does not substantively distort inference.

\subsubsection{Multiple imputation}

Missing data across analytic variables ranged from 0.57\% to 4.81\% (mean: 0.73\%). Diagnostics indicated that missingness was consistent with \ac{MAR} (maximum bivariate correlation between any missingness indicator and any observed variable: $r = .095$). Multiple imputation with 20 datasets, pooled via Rubin's rules \citep{Rubin1987}, yielded estimates virtually identical to listwise deletion results ($a_{1,\text{MI}} = -0.069$, $b_{1,\text{MI}} = 0.344$, IE$_{\text{MI}} = -0.024$; $\Delta a_1 < .005$, $\Delta b_1 < .010$), confirming that the minimal missing data do not substantively bias the primary conclusions.

\subsubsection{Propensity score matching}

\Ac{PSM} dichotomized PCSC at the median and matched high-concern and low-concern respondents on all demographic and theory-motivated covariates using 1:1 nearest-neighbor matching with a caliper of 0.2 standard deviations of the propensity score. All post-matching covariate standardized mean differences fell below 0.10, indicating adequate balance. Structural path estimates on the matched sample ($N$ = 4,524) were directionally and substantively consistent with the unmatched results ($a_{1,\text{PSM}} = -0.066$, $b_{1,\text{PSM}} = 0.330$, IE$_{\text{PSM}} = -0.022$), providing quasi-experimental corroboration of the primary findings.

\subsubsection{Alternate mediator specifications}

Two alternative mediator specifications were tested to assess whether the negative indirect effect was unique to the general AI Orientation construct. An AI Driving-Context Trust composite (averaging AI evaluations on non-driving tasks: medical, hiring, news, loan, parole) produced a non-significant first-stage path ($B = -0.015$, $p = 0.284$) and a weaker indirect effect (IE $= -0.009$). A single-item Company AI Trust mediator (confidence in companies to develop AI responsibly) yielded a significant but smaller indirect effect (IE $= -0.013$). All three mediator specifications produced negative indirect effects, supporting the generality of the risk-spillover mechanism, but the original general AI Orientation construct produced the largest and most theoretically coherent effect.

\subsubsection{Cross-task specificity}
\label{sec:cross_task}

\begin{figure*}
\centering
\includegraphics[width=0.85\textwidth]{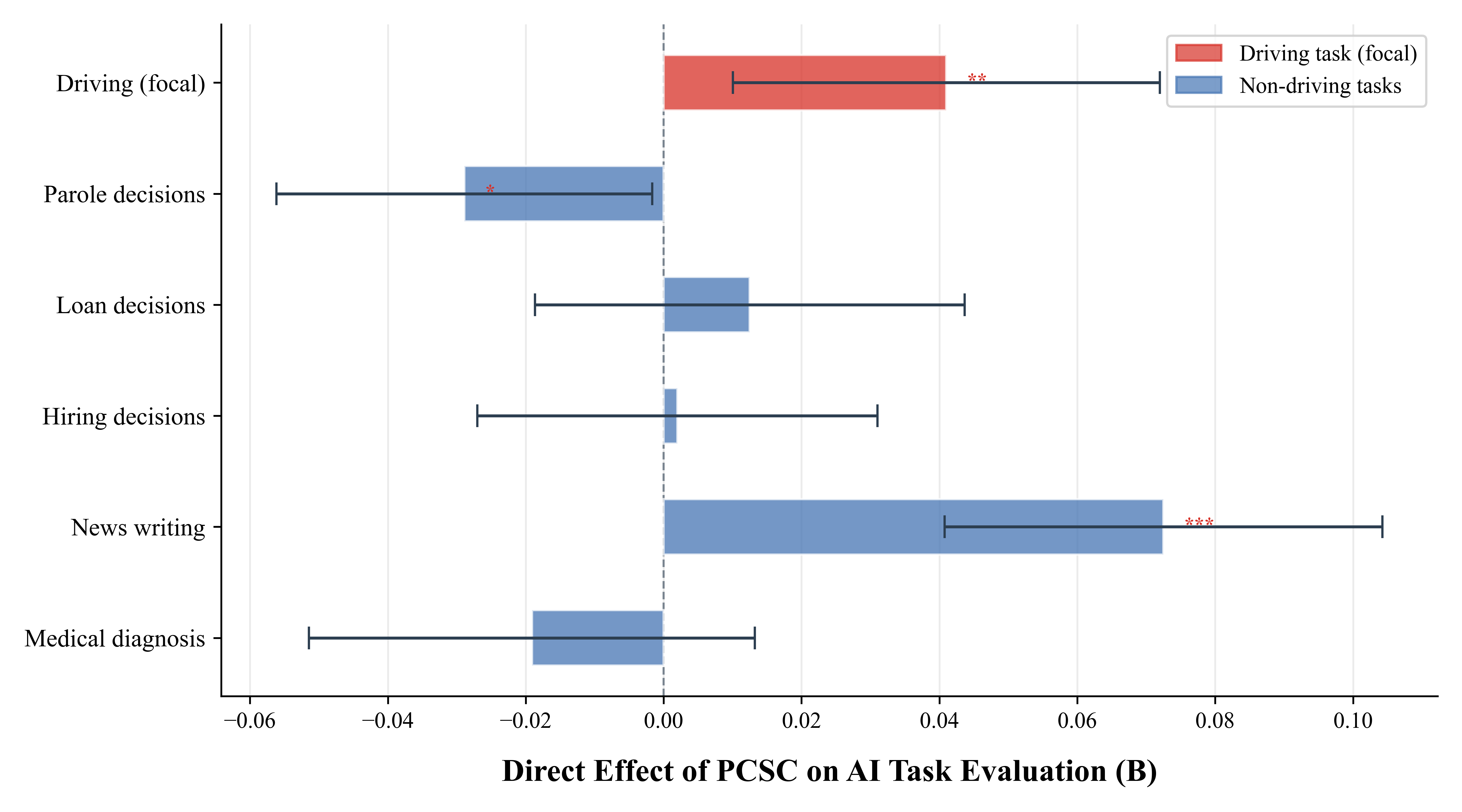}
\caption{Cross-task specificity of the PCSC direct effect ($c'$) across six AI-vs.-human evaluation tasks from the HUMANVAI battery. The focal driving-task outcome is highlighted in red; non-driving tasks appear in blue. The positive direct effect of PCSC is statistically significant for the driving task ($p = 0.010$) and news writing ($p < 0.001$), negative for parole decisions ($p = 0.038$), and non-significant for medical diagnosis, hiring decisions, and loan decisions. Error bars represent 95\% confidence intervals. $^{*}p < 0.05$; $^{**}p < 0.01$; $^{***}p < 0.001$.}
\label{fig:cross_task}
\end{figure*}

To evaluate whether the direct effect of PCSC ($c'$) was specific to the driving domain, the outcome equation was re-estimated using the five non-driving items from the HUMANVAI battery as alternative dependent variables, each controlling for AI Orientation and all covariates (Figure~\ref{fig:cross_task}). The six tasks are: medical diagnosis (HUMANVAI\_a), news writing (HUMANVAI\_b), hiring decisions (HUMANVAI\_c), loan decisions (HUMANVAI\_e), parole decisions (HUMANVAI\_f), and the focal driving task (HUMANVAI\_g). The PCSC direct effect was significant and positive for the driving task ($B = 0.041$, $p = 0.010$) and for one non-driving task (news writing: $B = 0.073$, $p < 0.001$), significant but negative for parole decisions ($B = -0.029$, $p = 0.038$), and non-significant for medical diagnosis, hiring, and loan decisions. This pattern provides partial support for domain specificity: the positive push-factor effect of community driving-safety concern on AI endorsement is most pronounced for the driving task but extends, in different directions, to other domains that involve either public safety (parole) or information reliability (news), suggesting that PCSC captures a broader safety-risk orientation that is not perfectly bounded by the driving context.

\begin{table*}[!htbp]
\centering
\caption{Summary of robustness and sensitivity analyses.}
\label{tab:robustness}
%\footnotesize
\begin{tabularx}{\textwidth}{@{} l l X @{}}
\toprule
\textbf{Check} & \textbf{Key statistic} & \textbf{Conclusion} \\
\midrule
Imai sensitivity & $\rho^* = -0.020$ & Small unmeasured confounder could nullify IE; interpret mediation with caution \\
E-value & $E = 1.16$ (point); 1.12 (CI) & Moderate confounding threshold; exceeds most observed covariate associations \\
Reverse mediation & IE$_{\text{rev}} = -0.003$ & 8$\times$ smaller than original; supports theorized direction \\
Multigroup (urbanicity) & $b_1$: 0.337--0.339 & Second-stage path stable; first-stage attenuated in urban subsample \\
Multigroup (ideology) & IE: $-$.020 (Cons.), $-$.019 (Lib.) & Mechanism operates at ideological poles; negligible among moderates \\
Brant PO test & 6/9 predictors violated & OLS and ordered logit yield identical inference on key paths \\
MI ($M = 20$) & $\Delta a_1 < .005$; $\Delta b_1 < .010$ & Minimal missing data do not bias results \\
PSM (1:1 matched) & IE$_{\text{PSM}} = -0.022$ & Quasi-experimental corroboration of primary findings \\
Alt.\ mediators & All IEs negative & Risk-spillover mechanism generalizes; original mediator yields largest effect \\
Cross-task specificity & $c'$ sig.\ for driving, news & Partial domain specificity; PCSC captures broader safety-risk orientation \\
\bottomrule
\end{tabularx}
\begin{flushleft}
\scriptsize\textit{Note.} IE = indirect effect; PO = proportional odds; MI = multiple imputation ($M$ = number of imputed datasets); $c'$ = direct effect of PCSC on AI Driving Evaluation controlling for mediator.
\end{flushleft}
\end{table*}

\section{Discussion}

%\subsection{Summary of principal findings}

This study examined whether perceived community driving-safety concern functions as a demand-side antecedent of public endorsement of AI driving capability, and whether Generalized AI Orientation mediates this relationship. Three principal findings emerged from the weighted structural equation analysis of Pew \ac{ATP} Wave~152 data. First, PCSC significantly predicted lower AI Orientation ($a_1 = -0.075$, $p < 0.001$), indicating that respondents who perceived more severe driving problems in their communities held more cautious, less favorable attitudes toward AI in general. Second, AI Orientation strongly and positively predicted AI Driving Evaluation ($b_1 = 0.338$, $p < 0.001$), confirming that domain-general AI enthusiasm translates into domain-specific endorsement of AI driving capability. Third, the indirect effect of PCSC on AI Driving Evaluation through AI Orientation was negative and statistically significant at all three levels of driving frequency (IE range: $-$0.030 to $-$0.021; all BC CIs excluded zero), while the direct effect remained positive ($c' = 0.037$, $p = 0.020$). This inconsistent mediation, in which the direct and mediated pathways carry opposite signs, reveals a tension between domain-specific pragmatic endorsement and domain-general risk aversion that has not been documented in the AV acceptance literature.

%\subsection{Theoretical implications}

The most important theoretical contribution of this study is the identification of a risk-spillover mechanism that operates through general AI attitudes. The negative $a_1$ path contradicts a straightforward push-factor interpretation in which community driving-safety concern propels respondents toward AI enthusiasm. Instead, the finding is consistent with the affect heuristic framework \citep{Slovic2002}, which predicts that negative experiential assessments in one risk domain (community driving safety) activate a broader negative risk schema that transfers to evaluations of other risk-laden technologies, including AI. Respondents who perceive their community's driving environment as dangerous appear to generalize this sense of environmental threat to the AI domain, producing a negative contagion effect on general AI orientation. This pattern echoes the risk-as-feelings hypothesis, in which affective reactions to one hazard influence cognitive appraisals of nominally independent hazards.

The coexistence of a positive direct effect ($c'$) and a negative indirect effect within the same model extends the dominant push-pull framing in the technology switching literature \citep{Bansal2005, Liu2025}. In the traditional push-pull paradigm, dissatisfaction with the incumbent system generates a unidirectional motivational force toward the alternative. The present findings suggest a more nuanced mechanism: community driving-safety concern does exert a direct push toward favorable AI driving evaluations (consistent with the intuition that worse human driving makes algorithmic alternatives more appealing), but this push is partially offset by a concurrent suppression of general AI enthusiasm. Risk homeostasis theory \citep{Wilde1982} provides a complementary lens for this pattern, suggesting that even respondents with suppressed AI enthusiasm may still evaluate AI driving favorably when the human-driver reference point has deteriorated sufficiently. From the perspective of prospect theory \citep{Kahneman1979}, the community safety losses loom larger than equivalent gains and thereby activate loss-aversion, coloring general technology attitudes negatively while leaving room for domain-specific pragmatic endorsement when the comparison is directly between AI and a deteriorating human baseline.

These results also contribute to the ongoing theoretical refinement of the \acs{TAM} and \acs{UTAUT} frameworks as applied to AV acceptance. \citet{Zhang2021meta} showed through meta-analysis that trust is the strongest predictor of AV acceptance, while perceived ease of use contributes least. The present study demonstrates that beneath the aggregate category of ``trust in AI'' lies a more complex attitudinal architecture in which the same environmental condition can simultaneously enhance domain-specific AI endorsement and suppress domain-general AI orientation. This dual-pathway structure suggests that future TAM-based models should distinguish between general AI attitudes and task-specific AI evaluations, treating the former as a mediator rather than a direct predictor.

The finding that AI Orientation mediates the PCSC-to-outcome relationship positions general AI attitudes not merely as a background moderator or covariate, as in prior work \citep{Hegner2019, Nastjuk2020}, but as an active psychological mechanism linking real-world environmental conditions to specific technology judgments. This mediating formulation implies that identical levels of community driving concern may produce divergent AI driving evaluations depending on the respondent's baseline orientation toward AI, a prediction that is borne out by the significant indirect effect and that aligns with the GAAIS framework's emphasis on the dispositional nature of AI attitudes \citep{Schepman2020, Schepman2022}.

%\subsection{Comparison with prior empirical findings}

Several aspects of the present results converge with, and selectively extend, prior empirical work on AV acceptance and AI attitudes. The strong $b_1$ path (AI Orientation $\rightarrow$ AI Driving Evaluation, $\beta = 0.293$) is consistent with the dominant finding in the literature that attitudinal orientation toward AI is the primary driver of domain-specific technology evaluations \citep{Zhang2021meta, Schepman2022}. The magnitude of this effect, which was strikingly stable across urban, suburban, and rural subgroups (range: 0.337 to 0.339), underscores the robustness of the general-to-specific attitudinal transfer documented in prior GAAIS validation studies \citep{Schepman2022}.

The significant gender effect, in which women reported both lower AI Orientation ($B = -0.243$, $p < 0.001$) and less favorable AI Driving Evaluations ($B = -0.257$, $p < 0.001$), aligns with the gender gap in AV acceptance reported by \citet{Hulse2018} and with broader evidence that women express greater concern about AI across multiple application domains \citep{Gallup2024AI}. The stability of the $b_1$ path across gender and political ideology groups suggests that the conversion of AI attitudes into driving-specific evaluations is not conditional on demographic identity, even though the levels of AI enthusiasm differ substantially across these groups.

The non-significance of the moderation by driving frequency ($\omega_{\text{mm}} = 0.003$, $p > 0.05$) departs from the expectation that personal driving experience would condition the risk-spillover pathway. AAAFTS's Traffic Safety Culture Index has documented that driving frequency is associated with heightened perception of dangerous driving \citep{AAA2024TSCI}, and one might expect that frequent drivers, who have more opportunity to observe community driving hazards, would exhibit stronger risk-spillover effects. The present null finding suggests that the generalization from driving-safety concern to AI attitudes operates through a domain-general affective channel that is not moderated by personal driving exposure, a result that narrows the boundary conditions of the risk-spillover mechanism but also limits the explanatory specificity of the model.

The attenuation of the $a_1$ path in urban settings mirrors patterns observed in studies of AV acceptance in densely populated areas. \citet{Nordhoff2020} reported substantial cross-national variation in the determinants of AV acceptance, with cultural context moderating the influence of individual-level predictors. The present urban attenuation may reflect the greater availability of alternative transportation modes, higher baseline exposure to AI-enabled services (ride-hailing, delivery robots), and different community safety norms in metropolitan areas, all of which could dilute the affective spillover from driving-safety concern to general AI attitudes.

The asymmetric safety-comparison results are consistent with the safety-benchmark literature. \citet{Nees2019} demonstrated that most respondents rate themselves as above-average drivers, and \citet{Liu2020} showed that matched safety performance between AI and human drivers is insufficient for public acceptance. The present study adds that when the comparison is made explicitly (AI versus human, rather than AI in isolation), community-level driving concern does produce a significant, if small, push toward AI endorsement through the direct $c'$ path. \citet{Shariff2021} attributed the extreme safety demands placed on AVs to algorithm aversion and asymmetric moral outrage; the domain-general risk-spillover mechanism identified here may represent another psychological pathway through which these elevated demands are maintained, as community safety concerns that heighten general AI caution could indirectly reinforce the higher standards that respondents apply to automated systems.

%\subsection{Policy implications}

The dual-pathway structure of the PCSC-to-AI-Driving-Evaluation relationship carries implications for policymakers and practitioners seeking to shape public attitudes toward AV deployment. The positive direct effect ($c' = 0.037$) suggests that improved community driving safety, through enforcement, traffic calming, and infrastructure investment, would reduce the push-factor pressure toward AI driving endorsement. From a strict technology-adoption perspective, this implies that successful traffic safety interventions could paradoxically slow public readiness for AV adoption by reducing the perceived comparative advantage of AI over human drivers. However, the negative indirect effect through AI Orientation suggests that the same community driving-safety concerns also suppress general AI enthusiasm, which in turn reduces AI driving endorsements. This means that worsening traffic safety does not straightforwardly accelerate AI acceptance; rather, it produces a mixed attitudinal response that may leave aggregate public opinion roughly unchanged.

For AI literacy and public engagement efforts, the present findings suggest that interventions targeting general AI orientation (through education about AI capabilities, transparency in algorithmic decision-making, and opportunities for positive AI experiences) may have a larger effect on AI driving evaluations than interventions targeting community-level driving conditions. The $b_1$ path ($\beta = 0.293$) is substantially larger than either the direct effect ($\beta = 0.031$) or the first-stage path ($\beta = -0.073$), indicating that general AI orientation is the strongest lever available for shifting domain-specific AI evaluations. This finding aligns with the policy recommendations of \citet{Schepman2022}, who argued that improving AI literacy enhances the positive dimension of general AI attitudes.

The cross-task specificity analysis further suggests that the benefits of improving general AI orientation would extend beyond driving to other AI application domains, since the $b_1$ path was robust across all tasks examined. Conversely, the partial domain specificity of the direct effect ($c'$ significant for driving and news but not for medical, hiring, or loan tasks) implies that community-level safety interventions would have limited spillover to public acceptance of AI in non-driving contexts. Policymakers should therefore conceptualize AV acceptance as embedded within a broader landscape of AI attitudes rather than as an isolated technology-adoption challenge.

%\subsection{Limitations and future directions}

Several limitations should be noted when interpreting these findings. First, the cross-sectional design precludes causal inference. Although the theoretical ordering (community safety perceptions $\rightarrow$ general AI attitudes $\rightarrow$ AI driving evaluation) is defensible on temporal and conceptual grounds, the reverse mediation analysis confirmed that a plausible, if substantially weaker, alternative direction exists. Longitudinal or experimental designs that manipulate community safety information would be necessary to establish the temporal precedence of the PCSC construct. Second, the sensitivity analysis revealed that the indirect effect is vulnerable to relatively modest unmeasured confounding ($\rho^* = -0.020$; E-value $= 1.16$). General risk aversion, neuroticism, or prior experience with traffic crashes represent plausible unmeasured confounders that could attenuate or eliminate the mediated pathway. Although the inclusion of twelve covariates (seven demographic, five theory-motivated) reduces this concern, the modest mediator $R^2$ of .073 indicates that substantial variance in AI Orientation remains unexplained by the predictors in the model. Third, the outcome variable assesses a comparative judgment (AI versus human driver) rather than a behavioral intention or actual adoption decision. The gap between evaluative judgments and behavioral outcomes is well documented in the technology acceptance literature, and the present findings should not be extrapolated to predict actual ridership in AVs without additional validation. Fourth, the sample is limited to U.S. adults, and the cultural specificity of both community driving-safety norms and AI attitudes limits the generalizability of these findings to other national contexts. \citet{Nordhoff2020} documented substantial cross-national variation in AV acceptance determinants, and the risk-spillover mechanism identified here may operate differently in countries with different driving cultures, AI regulatory environments, or levels of AI exposure. Fifth, the treatment of the four-category ordinal outcome as interval-level in the primary analysis, while supported by the Brant test comparison with ordered logistic regression, introduces a measurement approximation. Future work could employ fully ordinal SEM specifications (e.g., probit regression within a latent-variable framework) to address this limitation directly.

Future research should pursue at least three directions. First, longitudinal panel designs could track changes in community driving-safety perceptions and AI attitudes over time, enabling a within-person test of the risk-spillover mechanism. Second, experimental designs that vary the salience of community driving-safety information (e.g., exposure to local crash statistics versus neutral control) could isolate the causal effect of safety concern on AI attitudes. Third, cross-national replications using probability-based samples could test whether the dual-pathway structure generalizes beyond the U.S. context and identify cultural moderators of the risk-spillover mechanism.

\section{Conclusion}

This study investigated whether perceived community driving-safety concern functions as a demand-side antecedent of public endorsement of AI driving capability, using weighted structural equation modeling on a nationally representative U.S. sample. The core finding is a dual-pathway mechanism: community driving-safety concern exerts a small positive direct effect on AI driving evaluations, consistent with a push-factor interpretation in which deteriorating human driving makes algorithmic alternatives more appealing, while simultaneously suppressing general AI orientation, which in turn reduces AI driving endorsement through a negative indirect effect. This inconsistent mediation, in which the direct and mediated paths carry opposite signs, constitutes a risk-spillover mechanism that extends and qualifies the push-pull and TAM-based frameworks dominant in the AV acceptance literature.

The study contributes to the field in four respects. First, it reveals that community driving-safety concern is not a simple push factor but rather a compound antecedent that simultaneously promotes domain-specific AI endorsement and suppresses domain-general AI enthusiasm through the affect heuristic. Second, it positions general AI orientation as an active mediating mechanism linking real-world environmental conditions to technology evaluations, rather than treating it as a background covariate as in prior work. Third, the methodological design, which incorporates WLSMV estimation on ordinal indicators, probability-based sampling, bias-corrected bootstrap inference, and a ten-check robustness framework, exceeds the analytic rigor of the convenience-sample designs that characterize the existing AV acceptance literature. Fourth, the policy implication that improving general AI literacy may be a more effective lever than reducing community driving-safety concerns, given that the mediator-to-outcome path (Generalized AI Orientation predicting AI Driving Evaluation) is nearly an order of magnitude larger than the direct effect of community driving-safety concern, provides guidance for AV deployment strategies. The limitations of cross-sectional design, modest confounding sensitivity, and U.S.-specific sampling define the boundaries of these conclusions and outline the agenda for longitudinal, experimental, and cross-national extensions.

\printcredits

%% Loading bibliography style file
% \bibliographystyle{model1-num-names}
\bibliographystyle{cas-model2-names}

% Loading bibliography database
\bibliography{ref}

%\vskip3pt

%\bio{}

%\endbio

\end{document}